\documentclass[a4paper,fleqn,usenatbib]{mnras}
\usepackage[T1]{fontenc}
\usepackage{ae,aecompl}

\usepackage{graphicx}	% Including figure files
\usepackage{amsmath}	% Advanced maths commands
\usepackage{amssymb}	% Extra maths symbols
\usepackage{float}
\usepackage{natbib}
\usepackage{morefloats}
\usepackage{dcolumn}
\usepackage{amsmath}
\usepackage{latexsym}
\usepackage{longtable}
\usepackage{lipsum} % j
\begin{document}
\title[SN 2015as:  A low luminosity Type IIb supernova]
{SN 2015as:  A low luminosity Type IIb supernova without an early light curve peak}
\author[Gangopadhyay Anjasha et al.]
{Anjasha Gangopadhyay\thanks{e-mail :anjasha@aries.res.in, anjashagangopadhyay@gmail.com}$^{1,2}$, Kuntal Misra$^{1}$, A. Pastorello$^{3}$,
D.K.Sahu$^{4}$, 
\newauthor
L. Tomasella$^{3}$, L. Tartaglia$^{3}$, Mridweeka Singh$^{1,2}$, Raya Dastidar$^{1,5}$, 
\newauthor
S. Srivastav$^{4}$, P. Ochner$^{3}$,  Peter J. Brown$^{6}$, G. C. Anupama$^{4}$, 
\newauthor
S. Benetti$^{3}$, E. Cappellaro$^{3}$, Brajesh Kumar$^{4}$, Brijesh Kumar$^{1}$ and S. B. Pandey$^{1}$   \\
1. Aryabhatta Research Institute of observational sciencES, Manora Peak, Nainital 263 002 India \\
2. School of Studies in Physics and Astrophysics, Pandit Ravishankar Shukla University, Chattisgarh 492 010, India \\
3. INAF Osservatorio Astronomico di Padova, Vicolo dell'Osservatorio 5, 35122 Padova, Italy \\
4. Indian Institute of Astrophysics, Koramangala, Bangalore 560 034, India \\
5. Department of Physics $\&$ Astrophysics, University of Delhi, Delhi-110 007 \\
6. George P. and Cynthia Woods Mitchell Institute for Fundamental Physics $\&$ Astronomy, Texas A. $\&$ M. University, \\
Department of Physics and Astronomy, 4242 TAMU, College Station, TX77843, USA}

% These dates will be filled out by the publisher

% These dates will be filled out by the publisher
\date{Accepted XXX. Received YYY; in original form ZZZ}

% Enter the current year, for the copyright statements etc.
\pubyear{2016}

% Don't change these lines
%\begin{document}
\label{firstpage}
\pagerange{\pageref{firstpage}--\pageref{lastpage}}
\maketitle

% Abstract of the paper
\begin{abstract}
We present results of the photometric (from 3 to 509 days past explosion) and spectroscopic  (up to 230 days past explosion) monitoring campaign of the He-rich Type IIb supernova (SN) 2015as. The {\it (B-V)} colour evolution of SN 2015as  closely resemble those of SN 2008ax, suggesting that SN 2015as belongs to the SN IIb subgroup that does not show the early, short-duration photometric peak. The light curve of SN 2015as reaches the $B$-band maximum about 22 days after the explosion, at an absolute magnitude of -16.82 $\pm$ 0.18 mag. At $\sim$ 75 days after the explosion, its spectrum transitions from that of a SN II to a SN Ib. P~Cygni features due to He I lines appear at around 30 days after explosion, indicating that the progenitor of SN 2015as was partially stripped. For SN~2015as, we estimate a $^{56}$Ni mass of $\sim$ 0.08 M$_{\odot}$ and ejecta mass  of 1.1--2.2 M$_{\odot}$, which are similar to the values inferred for SN 2008ax. The quasi bolometric analytical light curve modelling suggests that the progenitor of SN 2015as has a modest mass ($\sim$ 0.1 M$_{\odot}$), a nearly-compact ($\sim$ 0.05$\times$10$^{13}$ cm) H envelope on top of a dense, compact ($\sim$ 2$\times$10$^{11}$ cm) and a more massive ($\sim$ 1.2  M$_{\odot}$) He core. The analysis of the nebular phase spectra indicates that $\sim$ 0.44 M$_{\odot}$ of O is ejected in the explosion. The intensity ratio of the [Ca II]/[O I] nebular lines favours either a main sequence progenitor mass of $\sim$ 15 M$_{\odot}$ or a Wolf Rayet star of 20 M$_{\odot}$. 
\end{abstract}

% Select between one and six entries from the list of approved keywords.
% Don't make up new ones.

%%%%%%%%%%%%%%%%%%%%%%%%%%%%%%%%%%%%%%%%%%%%%%%%%%

%%%%%%%%%%%%%%%%% BODY OF PAPER %%%%%%%%%%%%%%%%%%

\section{Introduction}
Supernovae of Type IIb (SNe IIb) are transitional objects whose spectra are dominated by hydrogen (H) lines at early phases, similar to Type II SNe. However, at later phases, their spectra share similarity with Type Ib SNe, being dominated by emerging helium (He) lines. The hydrogen-deficient core-collapse SNe are designated as Type Ib or Ic based on the presence or absence of helium features near about maximum light. SNe IIb, Ib and Ic are collectively known as stripped-envelope SNe. SNe IIb are limited in number, but they provide an interesting link between H-rich and H-poor core-collapse SNe. In a volume-limited sample consisting of 81 Type II SNe, \citet{2011MNRAS.412.1441L} found that Type IIb events are 11.9 $^{+3.9}_{-3.6}$$\%$. In a recent study by \cite{2017PASP..129e4201S}, stripped-envelope SNe constitute $\sim$ 10$\%$ among core-collapse SNe. Well studied SNe IIb include SN 1987K \citep{1988AJ.....96.1941F}, SN 1993J \citep{1993ApJ...415L.103F, 1994AJ....107..662A, 1994ApJ...426..334B,  1995A&AS..110..513B, 1996AJ....112..732R, 2000AJ....120.1499M, 2003astro.ph.10228F, 2015A&A...573A..12J}, SN 1996cb \citep{1999AJ....117..736Q, 2001astro.ph..6404D}, SN 2001ig \citep{2006MNRAS.369L..32R}, SN 2003bg \citep{2009ApJ...703.1624M, 2009ApJ...703.1612H}, SN 2008ax \citep{2008MNRAS.389..955P, 2011ApJ...739...41C, 2011MNRAS.413.2140T}, SN 2009mg \citep{2012MNRAS.424.1297O}, SN 2010as \citep{2014ApJ...792....7F}, SN 2011dh \citep{2011ApJ...742L..18A, 2013MNRAS.433....2S, 2015A&A...580A.142E}, SN 2011ei \citep{2013ApJ...767...71M}, SN 2011fu \citep{2013MNRAS.431..308K, 2015MNRAS.454...95M}, SN 2011hs \citep{2014MNRAS.439.1807B} and SN 2013df \citep{2014MNRAS.445.1647M}.
 
A fraction of SNe IIb (e.g. SNe 1993J, 2011fu and 2013df) show two peaks in their light curve, while other objects such as SNe 1996cb and 2008ax show only the broad maximum. The initial peak (observed in $\sim$ 50 $\%$ of SNe IIb, according to \cite{2016PhDT.......113M} ), is interpreted as a signature of the rapid cooling phase after the shock breakout, and lasts from seconds to days \citep{2016PhDT.......113M}. Thus, it can easily remain unobserved if the shock breakout occurs in a low density shell. The rapid decline phase is due to adiabatic cooling, and its duration depends primarily on the progenitor's size. The secondary peak is governed by the thermalisation of $\gamma$-rays and positrons during the radioactive decay of $^{56}$Ni $\rightarrow$ $^{56}$Co $\rightarrow$ $^{56}$Fe \citep{1970Ap&SS...8..457C,1970Ap&SS...8...20B}. The peak luminosities of most SNe IIb indicate that 0.02-0.1 M$_{\odot}$ of $^{56}$Ni is typically ejected during these explosions \citep{2013MNRAS.434.1098C, 2016MNRAS.458.2973P}. 

SNe IIb have unique spectroscopic features intermediate between those of Type II and Type I events. Because SN IIb spectra show H features at early phases which diminish over time while He features appear and strengthen, extended spectral sequences are essential to properly classify these transitional objects. Recent studies have claimed for the presence of some H in SNe Ib \citep{2005ApJ...631L.125A, 2007PASP..119..135P, 2016ApJ...820...75P}. A number of methods are used to draw a line between SNe IIb and Ib, including the estimate of the pseudo equivalent width of the H$\alpha$ line profile \citep{2016arXiv161207321L}, evaluating the ratio of the He I $\lambda$ 5876 to  H$\alpha$ equivalent widths \citep{2013ApJ...767...71M}, and through a comparison of the observed spectrum with Type IIb/Ib templates \citep{2016ApJ...820...75P}. 

In SNe IIb, strong stellar winds in single massive stars or mass outflows due to binary interaction may contribute to strip the outer stellar envelope of the progenitor. Type IIb can have compact (cIIb, with R $\sim$ $10^{11}$ cm) or extended (eIIb, with R $\sim$ $10^{13}$ cm) progenitors. \cite{2010ApJ...711L..40C} find similar properties of the radio light curve for IIb SNe that have been associated by other means with a compact/extended progenitor. They were the first to attempt a separation in extended and compact IIbs based on radio observations. Their observations indicated compact progenitors for SNe 1996cb, 2001ig and 2008ax, while those of SNe 1993J and 2001gd  are extended. However, more recent studies \citep{2014MNRAS.439.1807B, 2014MNRAS.440.1067R} claim that the inferred radio properties of a SN do not properly reflect the progenitor size.

Despite the small number of SNe IIb, direct detection for progenitors of four objects have been claimed. The progenitors were either Wolf-Rayet stars with M$_{ZAMS}$ = 10 - 28 M$_{\odot}$ in a binary system, like in the case of SN 2008ax \citep{2008MNRAS.389..955P, 2008MNRAS.391L...5C}, or more extended yellow supergiants with M$_{ZAMS}$ = 12 - 17 M$_{\odot}$ like in SNe 2011dh and 2013df \citep{2011ApJ...739L..37M, 2015MNRAS.454.2580M, 2014ApJ...793L..22F, 2011ApJ...741L..28V, 2013ApJ...772L..32V, 2014AJ....147...37V,  2012ApJ...757...31B}. For SN 1993J, the inferred progenitor was a K0Ia star in a binary system, with an early B-Type supergiant companion \citep{1993ApJ...415L.103F,1994AJ....107..662A, 2009Sci...324..486M}. Indirect indicators, such as the early time light curves, spectral studies of light echos and/or signatures of CSM-ejecta interaction, may also help constraining the progenitor mass like in the case of SNe 2001ig and 2013df \citep{2008Sci...320.1195K,2016ApJ...818..111K}. The early UV excess, for example, provides a method to estimate the progenitor radius  \citep{2015ApJ...803...40B}. In the case of SN 2001ig, both the radio light curves and the spectro-polarimetric observations support the binary progenitor \citep{2004MNRAS.349.1093R, 2007ApJ...671.1944M}, including a massive 10 - 18 M$_{\odot}$ Wolf-Rayet star \citep{2006MNRAS.369L..32R}. According to an alternative interpretation given by \cite{2006A&A...460L...5K}, the modulated late-time radio light curves would be a signature of a previous LBV-like pulsational phases.

In this paper, we present a photometric and spectroscopic study of the Type IIb SN 2015as. We provide information on the SN discovery and its host galaxy in Section \ref{1}. In Section \ref{2}, we describe the data acquisition and reduction procedures. In Section \ref{3}, we study the multi-band light curves, and compare the  colours, absolute light curves and bolometric evolution with those of other SNe IIb. The main explosion parameters are also computed through a basic analytical modelling of the bolometric light curve. The detailed description of the spectral evolution is presented in Section \ref{4}, while in Section \ref{5} a summary of the main results of this study is given.

\section{SN 2015as and its distance}
\label{1}
SN 2015as was discovered by Ken'ichi Nishimura\footnote{http://www.cbat.eps.harvard.edu/unconf/followups/J10081137+5150409.html} on three unfiltered CCD exposures taken on 2015 November 15.778 UT with a 35-cm reflector. The SN is located at RA = 10$^{h}$ 08$^{m}$ 11$^{s}$.37 and Decl. = +51$^{\circ}$ 50' 40.9'', which is 19".2 east and 2".9 north of the center of the SB(rs)d-Type host galaxy UGC 5460, which also hosted the Type IIn SN 2011ht \citep{2012ApJ...760...93H, 2012ApJ...751...92R, 2013ApJ...779L...8F, 2013MNRAS.431.2599M}. The unfiltered magnitude of the SN reported at the time of discovery was $\sim$16 mag. However, the SN was also visible on $r$-band images taken with the 182-cm Copernico Asiago Telescope on 2015 November 9, one week earlier than Nishimura's discovery during  routine monitoring of UGC 5460 \citep{2015ATel.8291....1T}. Using the spectrum taken on 2015 November 17.17 UT, it was classified as a Type II SN a couple of weeks after explosion. The spectrum, in fact, showed prominent Balmer lines, Ca H $\&$ K, Fe II  and also weak He I features \citep[][for a discussion on the explosion epoch, see Section 4]{2015ATel.8291....1T}.
At a redshift of {\it z} = 0.0036 \citep{1991rc3..book.....D}, the Virgo infall distance of UGC 5460 is 19.2 $\pm$ 1.4 Mpc  while the 3K CMB distance is 17.5 $\pm$ 1.2 Mpc (from NED where $H_{o}$ = 73.0 $\pm$ 5 km sec$^{-1}$ Mpc$^{-1}$ and using the flow model of \citet{2000ApJ...529..786M}). NED reports  the minimum and the maximum estimated redshift-independent Tully-Fisher distances to be 15.7 Mpc \citep{1985A&AS...59...43B} and 19.8 Mpc \citep{1988cng..book.....T}, respectively. The latter is in agreement with the Virgo infall distance (19.2 Mpc), which is the one adopted by \cite{2012ApJ...751...92R}, who found for SN 2011ht M$_{v}$ = -17 mag, which is typical of a core-collapse SNe. Consequently, we also adopt the same distance, 19.2 Mpc, for SN 2015as in this paper. 

The galactic reddening in the direction of SN 2015as is modest, {\it E(B-V)} = 0.008 $\pm$ 0.001 mag \citep{2011ApJ...737..103S}. The low-resolution spectra of SN 2015as (see Section 5) do not show the presence of the narrow interstellar Na I doublet (Na ID) absorption feature at the redshift of UGC 5460, suggesting that the extinction due to host galaxy dust is small. We, therefore, adopt {\it E(B-V)} = 0.008 $\pm$ 0.001 mag as the total reddening, corresponding to {\it A$_{V}$} = 0.025 mag  adopting the reddening law $R_V$ = 3.1 \citep{1973IAUS...52...31H}.
 
\section{Data acquisition and Reduction}
\label{2}
\subsection{Photometric Observations}
Photometric follow-up observations of SN 2015as were carried out a few days after the discovery using four ground-based optical telescopes. The observations were performed with the 104-cm Sampurnanand Telescope \citep[ST;][]{1999CSci...77..643G}, the 130-cm Devasthal Fast Optical Telescope  \citep[DFOT;][]{2012SPIE.8444E..1TS} located in ARIES India, the 182-cm Ekar Asiago Copernico Telescope, Italy\footnote{http://www.oapd.inaf.it/index.php/en/asiago-site/telescopes-and-instrumentations.html} and the 200-cm Himalayan Chandra Telescope  \citep[HCT;][]{2010ASInC...1..193P}, Indian Astronomical Observatory (IAO), India. The imaging observations were done using Johnson-Cousins-Bessell {\it B, V, R, I} and SDSS {\it u, g, r, i, z} filters. With the 104-cm ST, a 1K $\times$ 1K CCD camera was used, with a pixel scale of 0.37 arcsec pixel$^{-1}$, covering a 6.2 $\times$ 6.2 arcmin$^2$ field of view, while for the 130-cm DFOT, a 512 x 512 CCD was used with a plate scale of 0.63 arcsec pixel$^{-1}$ and covering 5.4 $\times$ 5.4 arcmin$^2$ area in the sky. The Asiago Faint Object Spectrograph and Camera (AFOSC) in Ekar Asiago telescope uses a 2K $\times$ 2K CCD with a plate scale of 0.26 arcsec pixel$^{-1}$. The 4K $\times$ 2K CCD on Himalayan Faint Object Spectrograph and Camera (HFOSC) was used for imaging with the HCT. The central 2K $\times$ 2K pixels on the chip were used for imaging which covers a 10 $\times$ 10 arcmin$^{2}$ region on the sky. 

Photometric monitoring of SN 2015as started on 2015 November 16 and continued up to 2017 March 28. Along with science frames, several bias and twilight flat frames were also collected. Bias, flat-field and cosmic ray corrected images were obtained and reduced using standard packages in IRAF\footnote{Image Reduction And Analysis Facility}. Instrumental magnitudes of the SN and field stars were obtained through aperture photometry (using DAOPHOT\footnote{Dominion Astrophysical Observatory + Photometry} package), with an optimal aperture which was usually three times the full width at half maximum (FWHM) of the stellar profile found using an aperture growth curve. PSF photometry was also performed, and using the field stars we obtained a well sampled PSF profile. This modelled PSF profile was used to measure the magnitudes of the SN and field stars. Even though the SN is located in the outskirts of its host galaxy--well isolated from significant contaminating sources--we preferred using the PSF magnitudes which minimizes any possible background contamination.

The Landolt standard fields \citep{1992AJ....104..340L} PG0918, PG0942, PG1323, PG1633 and SA98 were observed with HFOSC in the same nights as the SN field in the {\it B}, {\it V}, {\it R} and {\it I} bands. These standard fields allowed us to convert the instrumental magnitudes into apparent magnitudes. Observations were performed under good photometric conditions, with airmass varying between 1.1 - 1.5 and a typical seeing of 1.1 arcsec in the $V$-band. Instrumental and catalogue magnitudes of standard field stars were used to obtain zero points  and colour coefficients of the transformation equation, fitted using the least square regression technique, as described in \cite{1992JRASC..86...71S}. The average values of the colour coefficients and zero points obtained from the least square fits of 2016 April 8 and April 12 resulted in the following transformation equations 

{\it (V - v)} = 0.819 - 0.058 {\it (B - V)} + {\it k$_{v}$Q} 

{\it (B - b)} = 1.404 + 0.044 {\it (B - V)} + {\it k$_{b}$Q} 

{\it (I - i)} = 1.047 - 0.040 {\it (V - I)} + {\it k$_{i}$Q} 

\begin{figure}
	\begin{center}
		\hspace{-1.0cm}
		\includegraphics[scale=0.32]{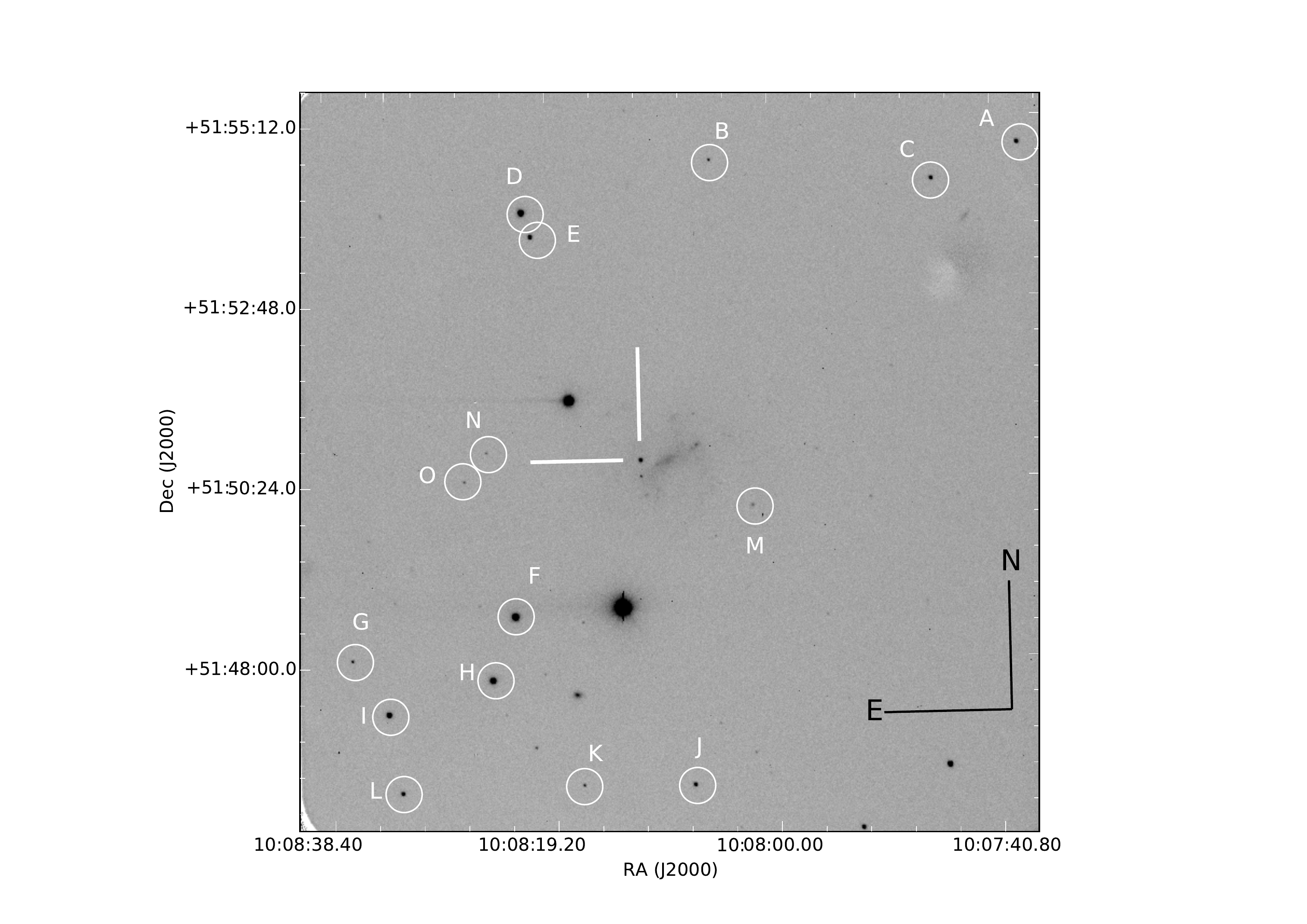}
	\end{center}
	\caption{SN 2015as and the local standard stars in the field of UGC~5460. $R$-band, 300 sec image obtained on 2016 February 24 with the 200 cm HCT.}
	\label{fig:calibimage}
\end{figure}

{\it (R - r)} = 0.823 - 0.070 {\it (V - R)} + {\it k$_{r}$Q} 

\noindent
where {\it Q} is the airmass and {\it k$_{\lambda}$} are the extinction coefficients. To correct for the atmospheric extinction, we used the site extinction values {\it k$_{v}$} = 0.12 $\pm$ 0.04, {\it k$_{b}$} = 0.21 $\pm$ 0.04, {\it k$_{i}$} = 0.05 $\pm$ 0.03 and {\it k$_{r}$} = 0.09 $\pm$ 0.04 mag airmass$^{-1}$ \citep{2008BASI...36..111S}. A root-mean-squared (rms) scatter between transformed and standard magnitude of Landolt stars was found to be between 0.02 -- 0.03 mag in the {\it BVRI} bands. Using these transformation equations, we calibrated 15 non-variable local standards in the SN field. These secondary standards are used to convert the SN instrumental magnitudes into apparent magnitudes in {\it BVRI} filters from EKAR/AFOSC, ST/1K x 1K CCD and DFOT/512 x 512 CCD. The local secondary standards in the SN field are marked in Fig \ref{fig:calibimage}, and their magnitudes are listed in Table \ref{tab:optical_observations}. For each night, precise zero points were determined using these secondary standards in order to correct for non-photometric conditions. The errors due to calibration and photometry were added in quadrature to estimate the final error in the SN magnitudes. The final SN magnitudes and their associated errors are listed in Table \ref{tab:observation_log}.  
%\newpage

The SN magnitudes in the $ugriz$ bands, obtained with the 182cm Ekar Asiago telescope were differentially calibrated using secondary standards obtained from SDSS DR12 \citep{2011AJ....142...72E}. We selected seven out of the 15 standard stars that are marked in Fig \ref{fig:calibimage} with known Sloan $ugriz$ magnitudes. They are listed in Table \ref{tab:optical_observations_SDSS}, while the Sloan-band SN magnitudes are listed in Table \ref{tab:observation_log}.

The {\it Swift} satellite  \citep{2004APS..APRS10001G,2004AIPC..727..637G}, equipped with the Ultra-Violet Optical Telescope  (UVOT; \citealp{2005SSRv..120...95R}), began observations of SN 2015as on 2015 November 16, and observed the SN field four times spaced two days apart. The {\it Swift} UVOT data were reduced using the procedure for the {\it Swift} Optical/Ultraviolet Supernova Archive (SOUSA; \citealp{2014Ap&SS.354...89B}). A pre-explosion image from 2012 June was used to subtract the contaminating host galaxy flux. The UVOT magnitudes are given in Table \ref{tab:UV_observations}. \\

\subsection{Spectroscopic Observations}
\label{Spectroscopy}
We obtained low and medium resolution optical spectra of SN 2015as at 24 epochs (from 2015 November 17 to 2016 June 23), using the 182-cm Ekar Asiago telescope equipped with AFOSC (Grisms Gr 4, VPH 6 and VPH 7) and HCT HFOSC (Grisms Gr 7 and Gr 8) (see Table \ref{tab:spec_observations}). A combination of grisms was frequently used to cover the entire optical region. The 2-d spectra were pre-processed using standard tasks in IRAF, as for the subsequent extraction and calibration of the 1-d spectra. FeNe, FeAr, HgCdNe arc lamps were used for wavelength calibration, whose accuracy was checked using night-sky emission lines. When appropriate, rigid wavelength shifts were applied. The spectroscopic standards GD71, BD+75d325 and Feige34 were observed for correcting the instrumental response and for flux calibration. The flux-calibrated spectra in the blue and the red regions were combined after scaling to get the final spectrum on a relative flux scale. The spectroscopic flux were checked with the photometric measurements at similar epochs and  appropriate scaling factors were applied. Finally, the spectra were corrected for redshift. 

\section{Photometric Results}
\label{3}
\subsection{Estimation of explosion epoch}
\begin{figure}
	\begin{center}
		\hspace{-0.6cm}
		\includegraphics[scale=0.4]{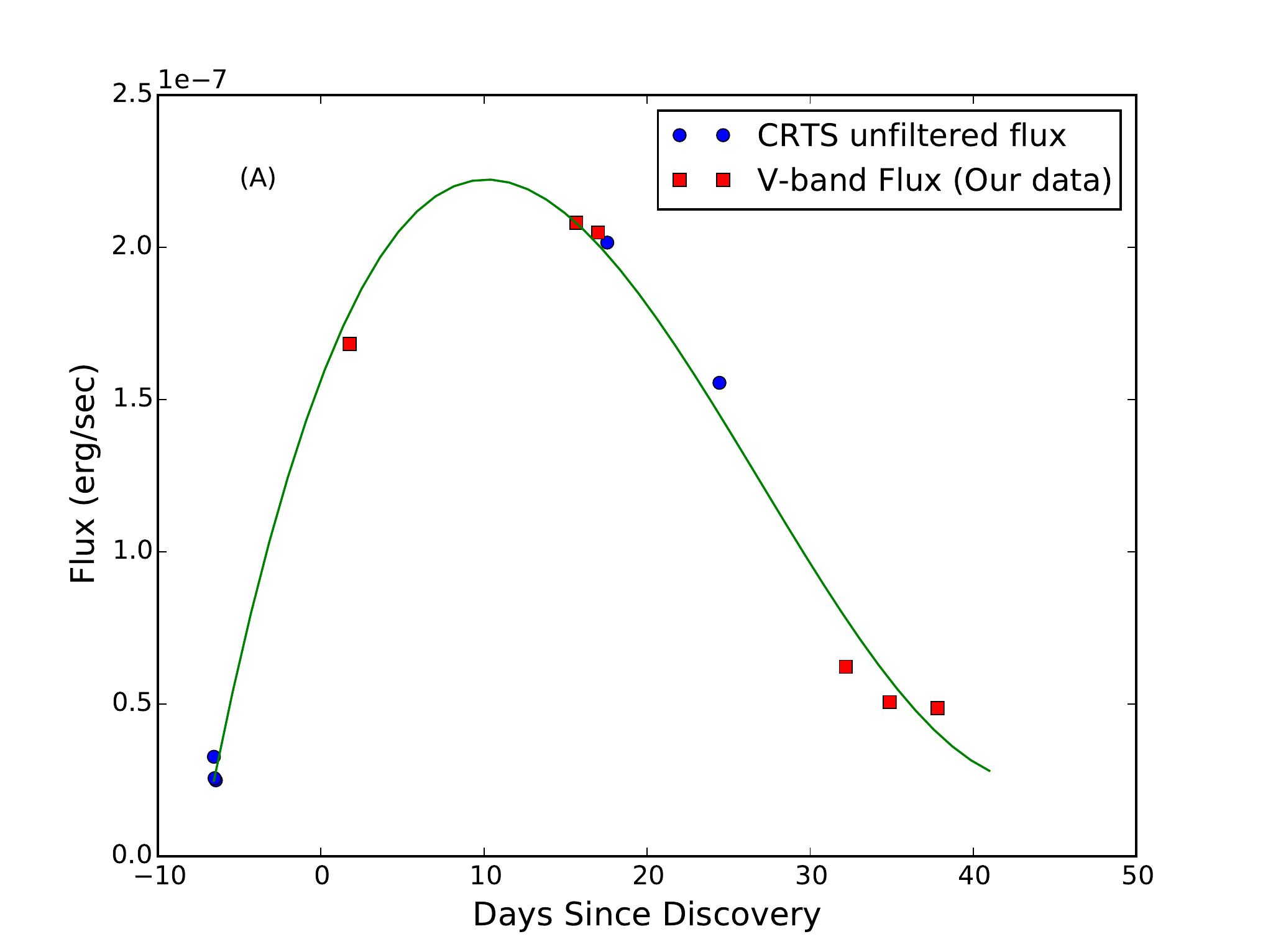} 
		\hspace{-0.8cm}
		\includegraphics[scale=0.4]{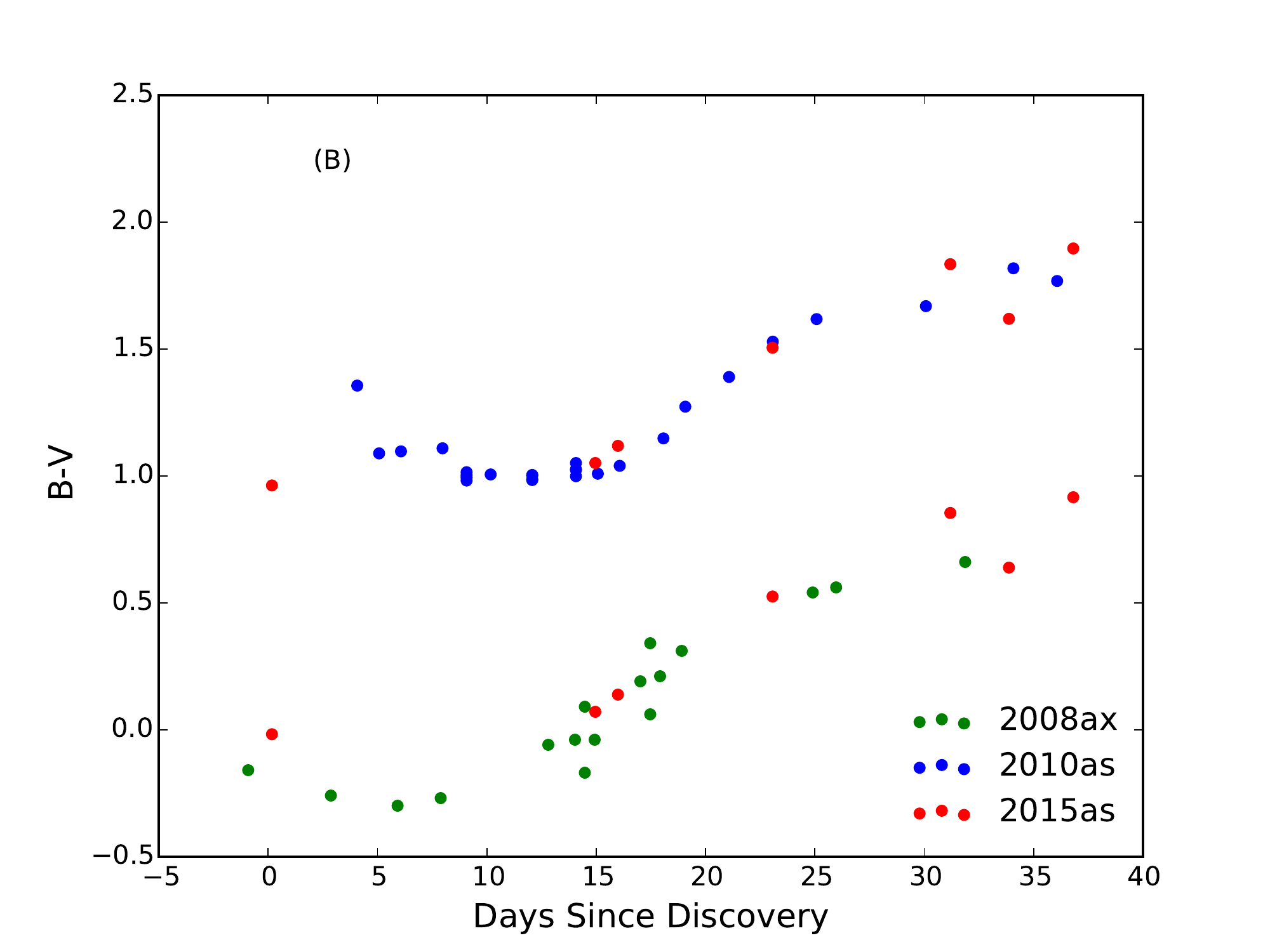} 
		\hspace{-0.8cm}
		\includegraphics[scale=0.4]{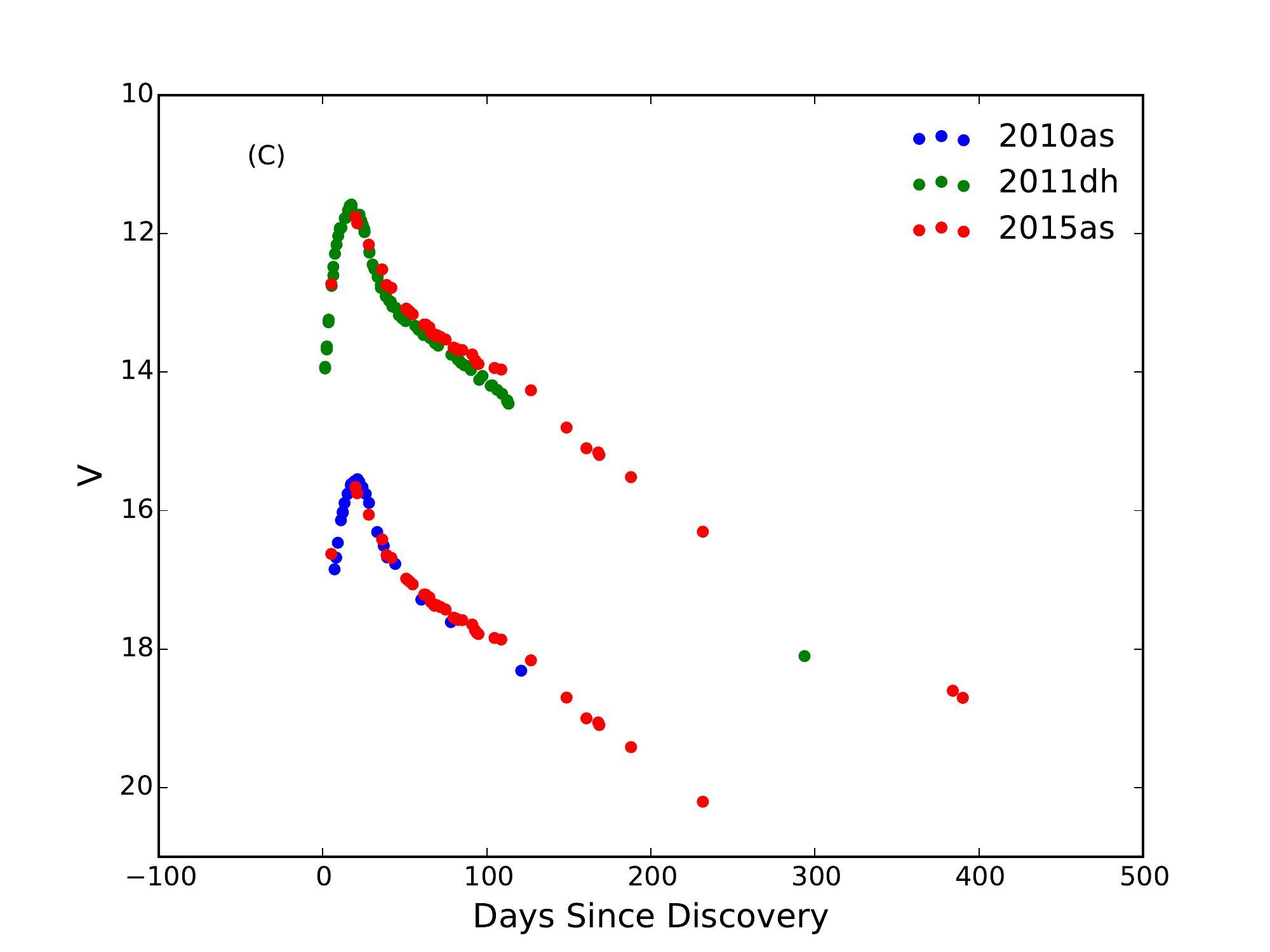} \\
	\end{center}
	\caption{(A) Plot showing parabolic fit of SN 2015as with others along with pre-explosion limit. (B) {\it B-V} colour comparison plot of SN 2015as with SNe 2008ax and 2010as. (C) {\it V}-band light curve comparison plot of SN 2015as with SNe 2008ax and 2010as.}
	\label{fig:comp_all}
\end{figure}

The SN was estimated to be 2-3 weeks past explosion from a preliminary spectroscopic analysis of the spectrum taken on 2015 November 17 \citep{2015ATel.8291....1T}. Although SN 2015as was discovered on 2015 November 15, a pre-discovery $r$-band detection was obtained on 2015 November 9 during routine monitoring of the galaxy UGC 5460 with the Asiago telescope. From this image, we measured a SN flux corresponding to $r(AB)$ = 17.59 $\pm$ 0.02 mag. We applied the SNID \citep{2011ascl.soft07001B} spectral cross-correlation tool to the spectra of SN 2015as taken on 2015 November 17 and December 03. In order to infer the SN explosion epoch, SNID cross-correlated the spectra of SN 2015as with other Type IIb templates, finding that the explosion occurred around 2015 November 5. The other tool GEneric cLAssification TOol (GELATO) \citep{2008A&A...488..383H}, which compares bins of the spectrum with those of template SNe, suggests the explosion occurred around 2015 November 2.   
Both results agree with the phase reported by \cite{2015ATel.8291....1T}.  SN 2015as was detected by the Catalina Real Time Transient Survey (CRTS) \citep{2009ApJ...696..870D} on 2015 November 8. Assuming the CRTS unfiltered magnitudes to be close to {\it V} band, we combine our {\it V}-band data
and CRTS data and perform a parabolic fit. CRTS applies transformations to the observed clear band magnitudes and the converted magnitudes are close to $V$ band\footnote{http://nessi.cacr.caltech.edu/DataRelease/FAQ.html}. The magnitudes are converted to flux and the fit is performed using the data upto nearly 38 days since discovery. The best-fit coefficients are used to find the roots of the equation i.e to  the value of time for which the flux equals zero. The best fit parabola is shown with the data in Fig. \ref{fig:comp_all} (A) From the best fit coefficients, the roots obtained are -8.01 days and +35 days. The first data point was on 2015 November 8 which is -6.4 days bfore discovery, so we estimate -8 days i.e 2015 November 6, as the explosion date.

\cite{2011ApJ...741...97D} estimate that for Type IIb and Ib SNe, explosion dates are usually $\sim$ 20 days prior to the $V$-band maximum. Moreover, SNe IIb exhibit the reddest {\it (B-V)} colour $\sim$ 40 days after explosion \citep{2008MNRAS.389..955P}. So, these two findings could serve as further tools to estimate the explosion epoch. We compared the evolution of {\it (B-V)} colours of SN 2015as with a few well-studied SNe IIb. We matched the maximum of {\it (B-V)} colour of SN 2015as and applied a shift in  phase to find the best matching SN from the SN IIb sample. The {\it (B-V)} colour evolution matching suggests {\it (B-V)} maximum occurred around 14 $\pm$ 1 days from discovery, thus the explosion epoch is again around 2015 November 6 ($\pm$ 1) days. This comparison gives an excellent match of SN 2015as with SNe 2008ax and 2010as (see Fig \ref{fig:comp_all} (B) ) which gives us an explosion epoch around 2015 November 6 ($\pm$ 1). Since the overall {\it V}-band light curve of SN 2015as is similar to that of SN 2011dh, we matched the light curve of SN 2015as with SNe 2010as and 2011dh by applying appropriate shifts in magnitude and phase (see Fig \ref{fig:comp_all} (C) ). This cross-matching of the light curves suggests $V$ max occurred on 15 $\pm$ 1 days from discovery, thus the explosion epoch is again around 2015 November 6 ($\pm$ 1) days. Accounting for all the above arguments, we finally estimate the explosion date of SN 2015as as 2015 November 6 ($\pm$2) (JD = 2457332.5) and use that date throughout the paper.

\subsection{Primary Light Curve Features}
\begin{figure}
	\begin{center}
		\vspace{-1.0cm}
		\hspace{-1.2cm}
		\includegraphics[scale=0.345]{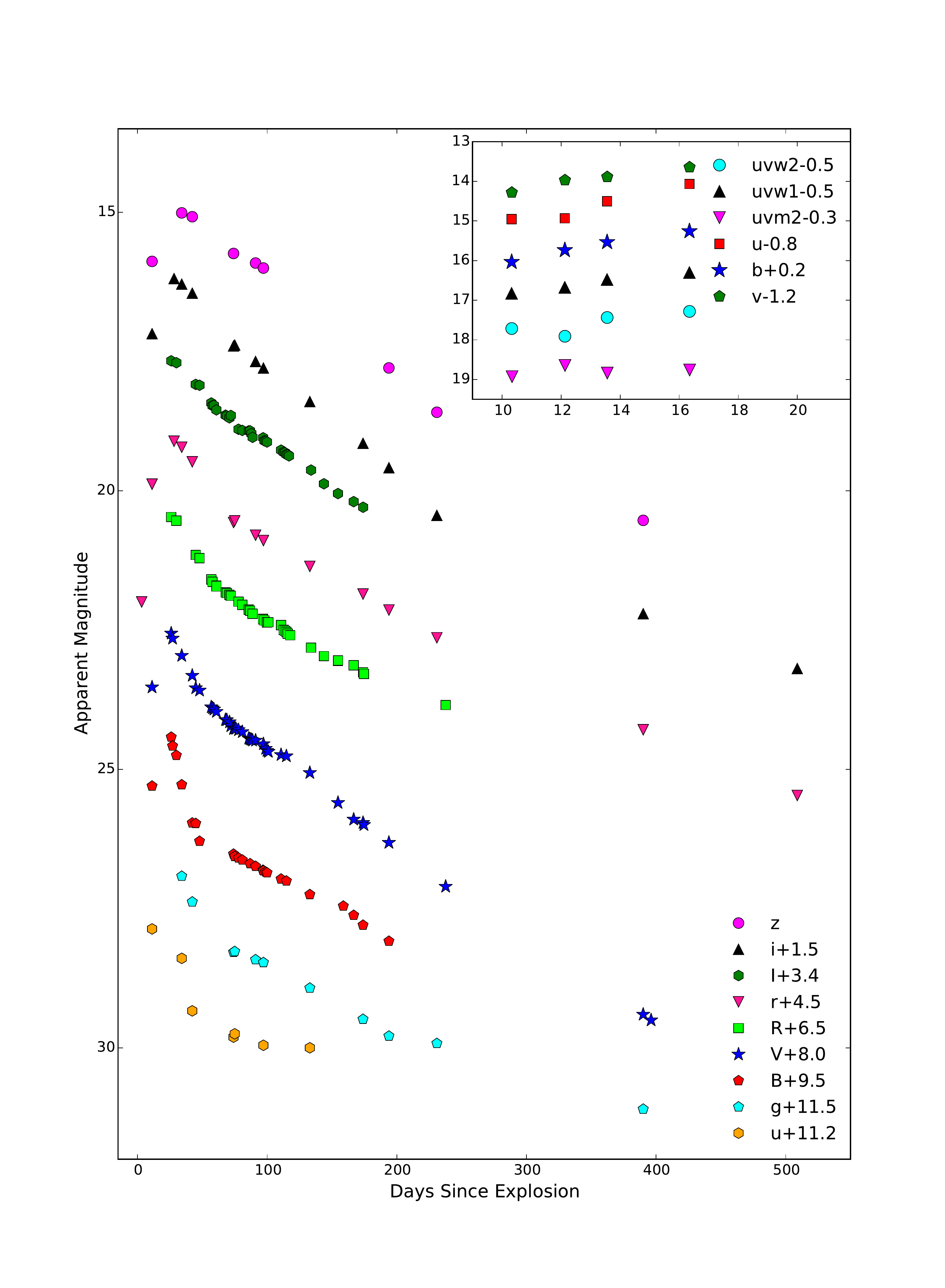}
	\end{center}
	\caption{$BVRI/ugriz$ light curve evolution of SN 2015as. Inset plot shows the light curve evolution of the UVOT filters. Explosion epoch is taken as 2015 November 6 (JD=2457332.5). The light curves are shifted arbitrarily for clarity.}
	\label{fig:plotfinal}
\end{figure}

The complete multi-band ({\it BVRIugriz} filters) light curves of SN 2015as from 3 to 509 days after explosion, and the early {\it Swift} UVOT observations are presented in Fig \ref{fig:plotfinal}. 
Several SNe IIb exhibit an early, fast evolving light curve peak. This early peak is not seen in SN 2015as (in analogy with SN 2008ax) which may be either due to the late discovery of the SN or its effective absence. Our observations started 3.1 days after the explosion. The peak magnitudes and epoch of maximum in each filter was estimated by a cubic spline fit.  The errors reported are obtained from the interpolated measurements around the peak. The {\it B} band maximum occurred 22.0 $\pm$ 0.3 days after the explosion, at an apparent magnitude of 14.74 $\pm$ 0.02 mag, while the $V$-band maximum occurred 4.0 days after the $B$-band maximum, at an apparent magnitude of 14.63 $\pm$ 0.03. The $r$ and $i$ bands peaked 2.7 days and 5.6 days after the $B$- band maximum, with apparent magnitudes of 14.48 $\pm$ 0.01 and 14.68 $\pm$ 0.01, respectively. The rise times to the light curve maxima in each band, their JDs at peak and their peak magnitudes  are tabulated in Table \ref{tab:photometric_properties}. The $B$-band maximum light is reached 22 days after the first detection, and there is a gradual delay in reaching the maximum light in the redder pass bands. 

\begin{figure}
	\begin{center}
		%\vspace{-1.0cm}
		\includegraphics[scale=0.34]{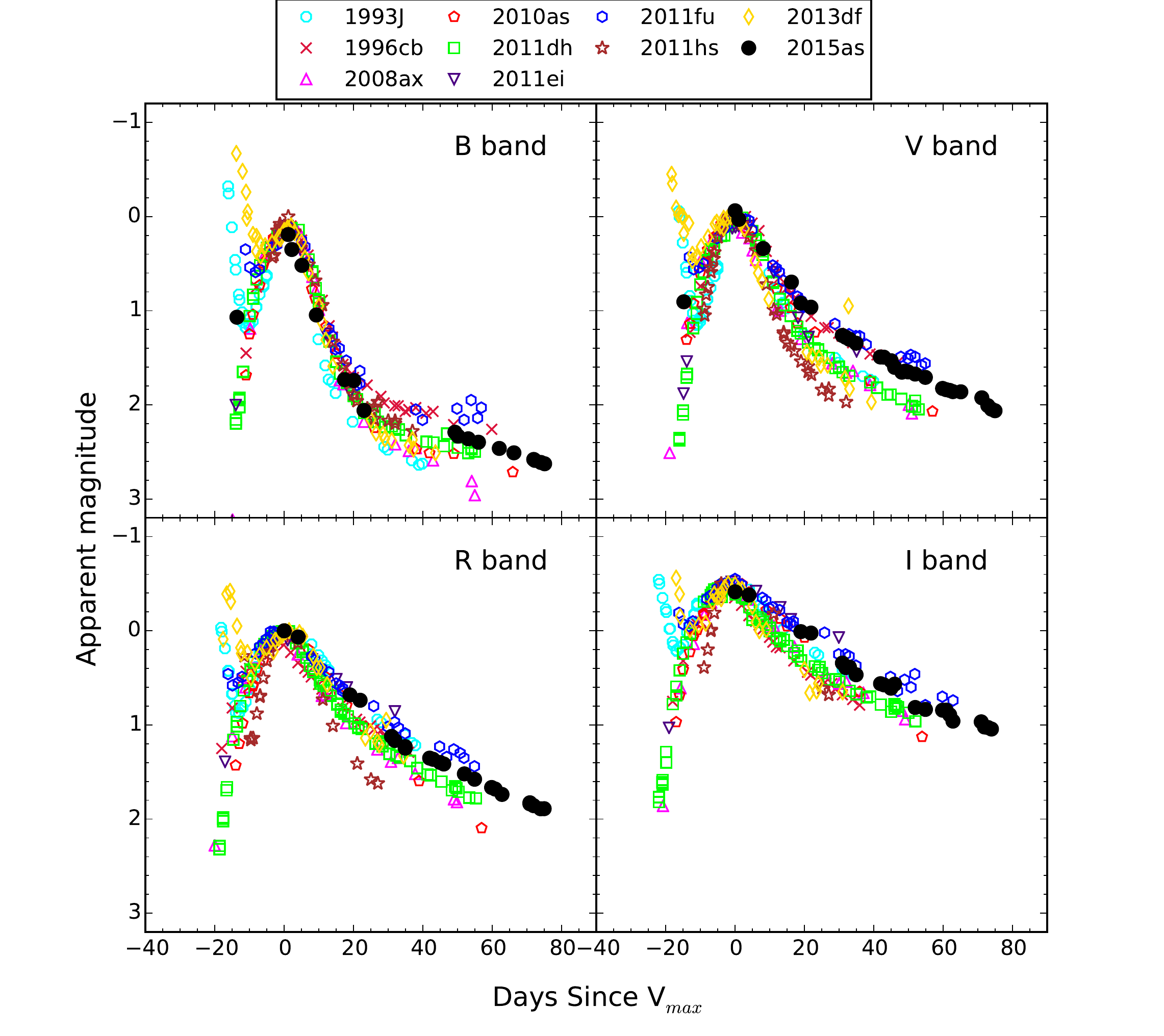}
	\end{center}
	\caption{A comparison of light curve evolution of SN 2015as and other SNe IIb in the different bands up to 100 days post explosion. The observed magnitudes are normalised with respect to their peak magnitudes and a shift in time is applied to match the epoch of maximum.}
	\label{fig:lightcurvecompearly}
\end{figure}

\begin{figure}
	\begin{center}
		%\vspace{-1.0cm}
		\includegraphics[scale=0.34]{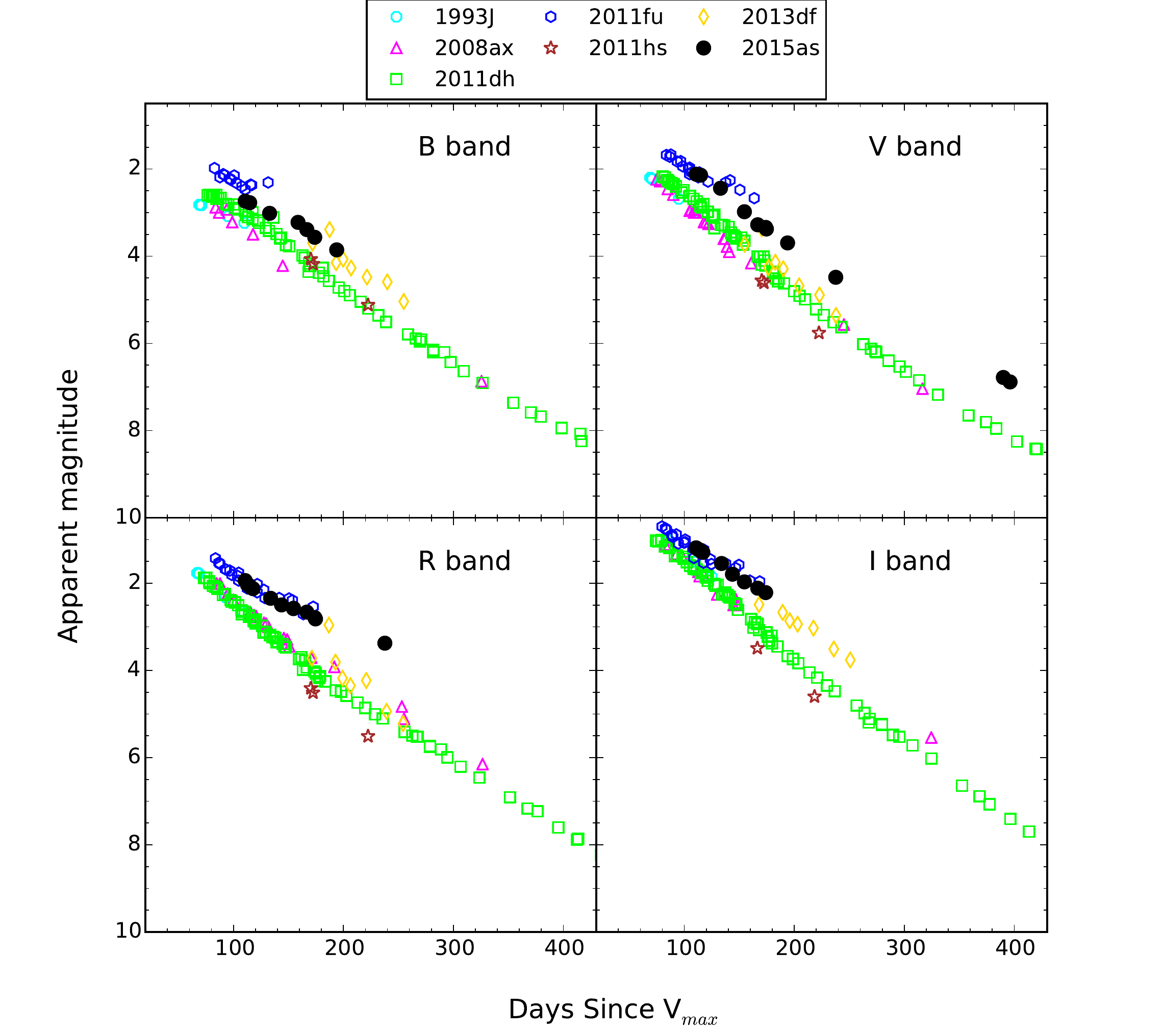}
	\end{center}
	\caption{Late time (80-400 days post explosion) light curve evolution of SN 2015as and other SNe IIb. The observed magnitudes are normalised with respect to their peak magnitudes and shifted in time to match the epoch of maximum.}
	\label{fig:lightcurvecomplate}
\end{figure}

The light curve features of SN 2015as are compared with those of other well-studied SNe IIb (cf. Fig \ref{fig:lightcurvecompearly} and Fig 
\ref{fig:lightcurvecomplate}). We construct a sample of ten SNe IIb (SN 1993J \citep{1994AJ....107.1022R, 1996AJ....112..732R}; SN 1996cb \citep{1999AJ....117..736Q}; SN 2003bg \citep{2009ApJ...703.1624M}; SN 2008ax \citep{2008MNRAS.389..955P, 2011MNRAS.413.2140T}; SN 2010as \citep{2014ApJ...792....7F}; SN 2011hs \citep{2014MNRAS.439.1807B}; SN 2011dh \citep{2013MNRAS.433....2S}; SN 2011ei \citep{2013ApJ...767...71M}; SN 2011fu \citep{2013MNRAS.431..308K, 2015MNRAS.454...95M} and SN 2013df \citep{2014MNRAS.445.1647M}) from the literature. The parameters of the comparison sample are listed in Table \ref{tab:photometric_parameters_different_SNe}.
 
SN 2015as shows comparatively larger rise times than other SNe IIb (Table \ref{tab:rise times}). Fig \ref{fig:lightcurvecompearly} shows the light curve evolution of our SN sample up to 100 days. Their observed magnitudes are normalised with respect to their peak magnitudes, and a shift in time is applied to match the time of maximum. At early epochs (pre-max phases), the $B$ and $V$ light curves of SN 2015as clearly resemble those of SNe 2011dh and 1993J. The decline in magnitude during the first 15 days after maximum  are $\triangle m_{15}$($B$) = 1.51 $\pm$ 0.16 and $\triangle m_{15}$($V$) = 0.68 $\pm$ 0.03 mag for the $B$ and $V$-bands, respectively. The $V$-band decline value is smaller than in other SNe IIb (see Table \ref{tab:photometric_parameters_different_SNe}). Our estimate of $\triangle m_{15}$($V$) for SN 2015as is also smaller than the average value for Type Ib/Ic SNe \citep[$\triangle m_{15}$($V$) = 0.8 $\pm$ 0.1 mag;][]{2011ApJ...741...97D}. Between 50 and 100 days, SN 2015as declines with rates of 1.10 $\pm$ 0.03, 1.85 $\pm$ 0.06, 1.87 $\pm$ 0.08 and 1.73 $\pm$ 0.01 mag (100 days)$^{-1}$ in the {\it B, V, R} and $I$-bands respectively. The decay rate of SN 2015as in $B$-band closely matches that of SN 2011dh whereas it is slower than SN 2011fu. The $V$-band decline of SN 2015as is faster than SNe 2011dh and 2011fu, whereas the $R$-band and the $I$-band decay rates of SN 2015as are slower than SNe 2011dh and 2011fu and are faster than SN 1993J (see Table \ref{tab:comp_decay_rate}). 

At 100-300 days, the slope of the light curve changes. A steepening is noticed in the $B$-band as compared with other SNe IIb, with the decay rate being 1.30 $\pm$ 0.08 mag (100 days)$^{-1}$, while a softening is noticed in the $V$-band decline with a rate of 1.66 $\pm$ 0.04 mag (100 days)$^{-1}$. The $B$-band steepening may be due to early dust formation. Dust formation always shifts the spectral energy distribution to the redder wavelengths, so early steepening or the fast decay of $B$-band could be an indication of early circumstellar interaction leading to dust formation. The $R$ and $I$-band light curves show decay rates of 1.13 $\pm$ 0.04 and 1.67 $\pm$ 0.03 mag (100 days)$^{-1}$ respectively. The late-time ($\geq$ 100 days) evolution of SN 2015as is shown in Fig \ref{fig:lightcurvecomplate}, along with those of other SNe IIb. The decay rates at comparable epochs for the IIb SN sample are listed in Table \ref{tab:comp_decay_rate}. The post-maximum light curves of SNe IIb are steeper in the blue bands, with a gradual flattening observed in red bands. The late-time decay rates of SNe IIb usually exceed the  $^{56}$Co $\rightarrow$ $^{56}$Fe decay rate (0.98 mag (100 days)$^{-1}$). This is indicative of incomplete gamma-ray trapping and/or lower positron deposition and is typical in SE SNe. The $B$-band light curve of SN 2015as shows a sort of plateau between 70-100 days after explosion, similar to SNe 1993J \citep{1994MNRAS.266L..27L} and 2011fu \citep{2013MNRAS.431..308K} at comparable epochs.

\subsection{Colour Curves}
\begin{figure}
	\begin{center}
		\hspace{-1.0cm}
		\includegraphics[scale=0.40]{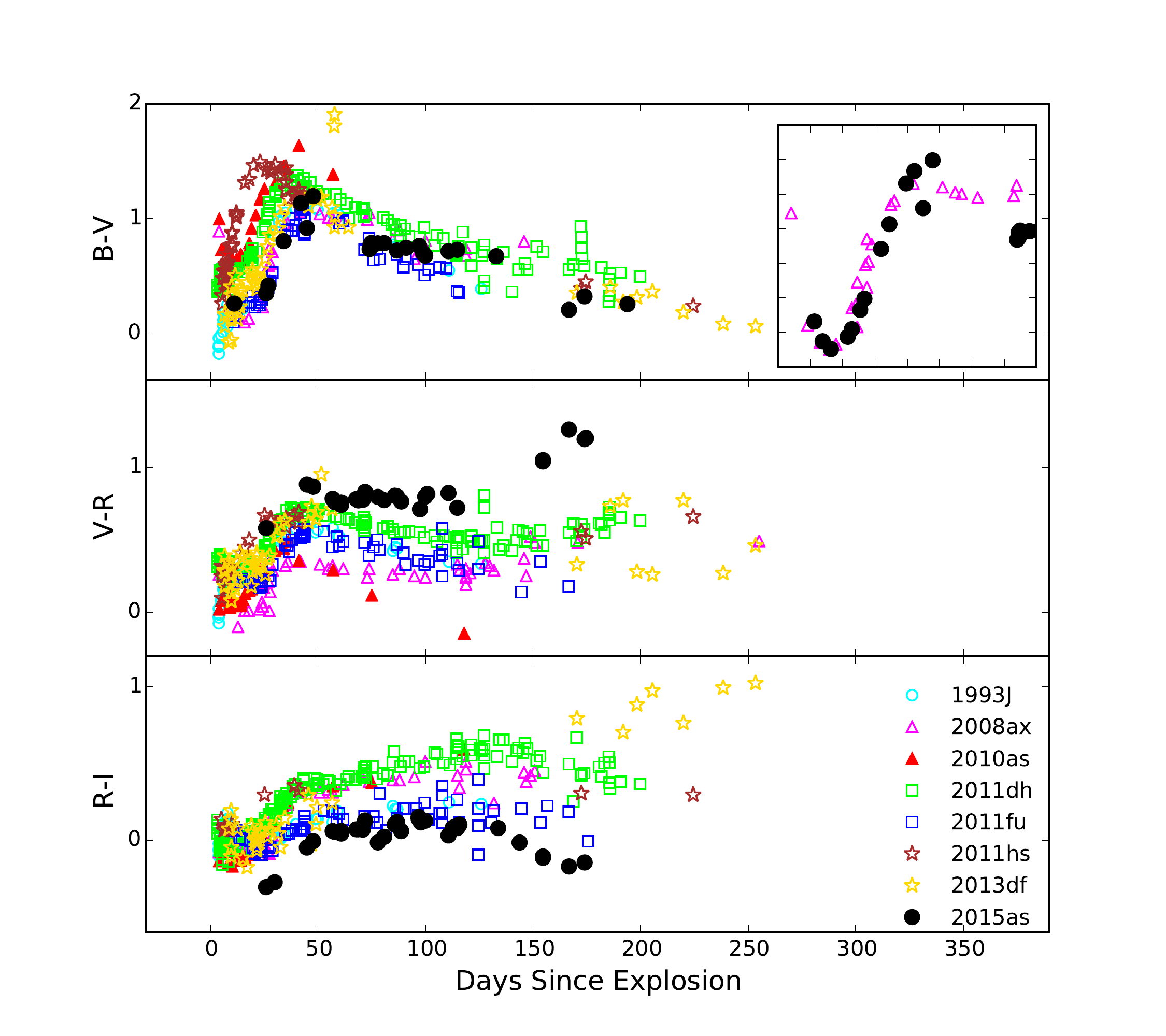}
	\end{center}
	\caption{A comparison of the {\it (B-V), (V-R)} and {\it (R-I)} colours of SN 2015as and other representative SNe IIb. The colours are corrected for the reddening values given in Table \ref{tab:photometric_parameters_different_SNe}. The inset in the topmost panel shows the {\it (B-V)} colours of SNe 2008ax and 2015as upto $\sim$ 80 days post explosion.}
	\label{fig:colorcurves}
\end{figure}

The colour evolution of SN 2015as is shown in Fig \ref{fig:colorcurves}. SNe 1993J, 2008ax, 2010as, 2011dh, 2011ei, 2011fu, 2011hs and 2013df are also shown as comparisons. All colour curves have been corrected for the reddening values given in Table \ref{tab:photometric_parameters_different_SNe}. The $(B-V)$ colours of SN 2015as become bluer from 11 to 25 days after explosion starting from $(B-V)$ $\sim$ 0.26 mag, then becomes redder from  about 25 to 50 days after explosion, reaching $(B-V)$ $\sim$ 1.19 mag, and evolves towards a blue colour reaching $(B-V)$ $\sim$ 0.26 mag at $\sim$190 days. During the first 10 to 20 days prior to maximum light, the colour of single-peaked SNe IIb like SNe 2008ax, 2010as shows an early red to blue transition that is not detected for double peaked SNe like SNe 1993J, 2011fu, 2013df etc \citep{2016PhDT.......113M}. SN 2015as shows such early red to blue $(B-V)$ colour transition. The $(V-R)$ moves to redder colours until $\sim$40 days after explosion (reaching 0.86  mag), then becomes bluer from 50 to 90 days. Beyond 170 days, the $(V-R)$ colour grows, and reaches a value of 1.75 mag which is redder than other SNe in the sample. The $(V-I)$ colour becomes redder from 25 to 70 days post explosion. From 70 to 154 days post explosion, it remains constant and becomes redder again reaching a value of 1.04 mag at 160 days. At this stage, the outer ejecta becomes optically thin. The redder colour of SN 2015as suggests that the SN is intrinsically redder.  We see that NaID due to the Milky Way and the host galaxy are not prominent, hence suggesting very modest extinction.

\subsection{Absolute Magnitude}
The absolute magnitudes of the SN sample are estimated using the distance and reddening values given in Table \ref{tab:photometric_parameters_different_SNe}. The distances of all the SNe are calculated assuming $H_{0}$ = 73.0 $\pm$ 5.0 km sec$^{-1}$ Mpc$^{-1}$. The $V$ peak band absolute magnitude of SN 2015as is -16.82 $\pm$ 0.18 mag, obtained by adopting the reddening and distance values given in Section 2. The peak absolute magnitude of SN 2015as is nearly 0.5 mag fainter than the mean absolute magnitude (-17.40 $\pm$ 0.10 mag; \citealp{2006AJ....131.2233R}) of SNe IIb  and 1.25 mag fainter than average peak absolute magnitude of Type Ib/c SNe. It is fainter than SN 1993J (-17.57 $\pm$ 0.24 mag ; \cite{1994AJ....107.1022R}), SN 2008ax (-17.61 $\pm$ 0.43 mag ; \cite{2016PhDT.......113M}), SN 2010as (-18.01 ; \cite{2016PhDT.......113M}), SN 2011dh (-17.12 $\pm$ 0.18 mag; \cite{ 2013MNRAS.433....2S}), SN 2011fu (-18.50 $\pm$ 0.24 mag; \cite{2013MNRAS.431..308K}), while it is nearly 0.5 mag brighter than SN 1996cb (-16.22 mag; \cite{1999AJ....117..736Q}) and SN 2011ei (-16.0 mag; \cite{2013ApJ...767...71M}). SN 2015as is of comparable brightness to  SN 2003bg (-16.95 mag; \cite{2009ApJ...703.1624M}), SN 2011hs (-16.59 mag; \cite{2014MNRAS.439.1807B}) and SN 2013df (-16.85 $\pm$ 0.08 mag; \cite{2014MNRAS.445.1647M}). SNe IIb show a moderate range of $V$-band peak absolute magnitudes (-18 to -16 mag) with SN 2015as belonging to the faint SNe IIb subgroup.

\subsection{Bolometric Light Curve} 
\begin{figure}
	\begin{center}
		%\vspace{-1.0cm}
		\includegraphics[scale=0.48]{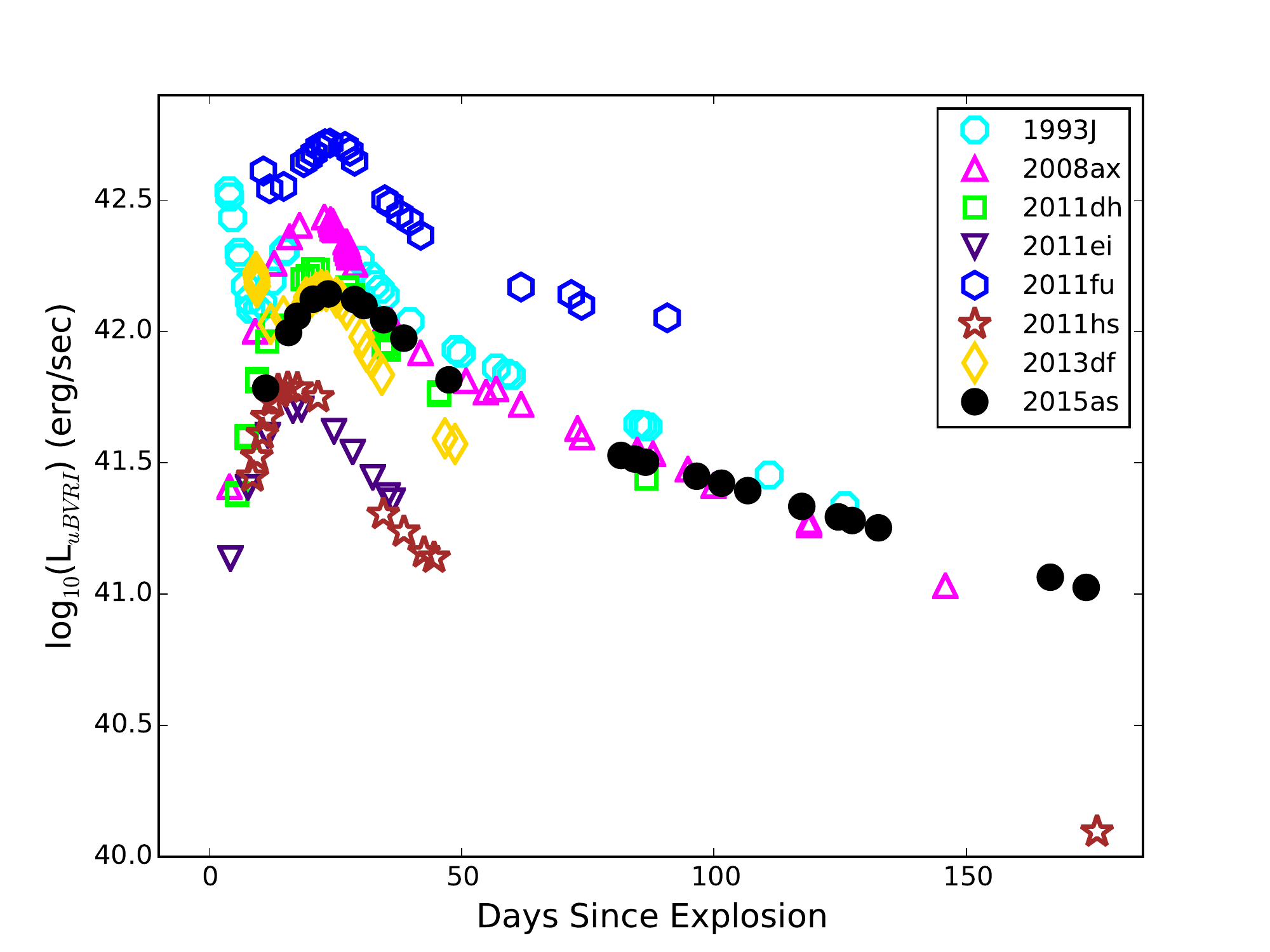}
	\end{center}
	\caption{Pseudo-bolometric light curves of our sample of SNe IIb. The distance and reddening values adopted for estimating their bolometric luminosities are given in Table \ref{tab:photometric_parameters_different_SNe}.}
	\label{fig:bol}
\end{figure}

In order to construct the bolometric light curve, the reddening-corrected magnitudes were converted to fluxes using the flux zero point values provided by \cite{1998A&A...333..231B} and the $u$ band zero point was taken from \cite{1996AJ....111.1748F}. The bolometric flux of SN 2015as at different epoch is estimated by integrating %(from 3000 to 9500 \AA) 
the  monochromatic fluxes in the $uBVRI$ bands using the trapezoidal rule. A distance of 19.2 $\pm$ 1.4 Mpc is adopted to convert the flux to bolometric luminosity. Our $u$-band observations end at 132 days after explosion.  Assuming a constant $(u-B)$ colour at very late phases, we estimate the $u$-band magnitudes up to 173 days using the constant $(u-B)$ colour equal to the last available one (132 days). We construct the pseudo-bolometric light curve using $uBVRI$ observations up to 173 days after explosion. To account for the missing observed flux on some nights, interpolation of the light curves is done. The pseudo-bolometric light curve of SN 2015as is shown along with those of other SNe IIb in Fig \ref{fig:bol}. The pseudo-bolometric light curves of SN 1993J \citep{1994AJ....107.1022R}, SN 2008ax \citep{2008MNRAS.389..955P,2011MNRAS.413.2140T}, SN 2011dh \citep{2013MNRAS.433....2S}, SN 2011ei \citep{2013ApJ...767...71M}, SN 2011fu \citep{2013MNRAS.431..308K}, SN 2011hs \citep{2014MNRAS.439.1807B} and SN 2013df \citep{2014MNRAS.445.1647M} in the optical domain are constructed with the same method as SN 2015as. The distance and reddening values are given in Table \ref{tab:photometric_parameters_different_SNe}. SN 2015as is fainter than most SNe IIb, except SNe 2011hs and 2011ei. In fact, the brightest SN IIb in our sample, SN 2011fu, has a peak luminosity of 5.2 x 10$^{42}$ erg sec$^{-1}$ whereas SN 2015as has a peak luminosity which is a factor 3.2 lower (1.5 x 10$^{42}$ erg sec$^{-1}$).   

Assuming that the radioactive decay of $^{56}$Ni $\rightarrow$ $^{56}$Co $\rightarrow$ $^{56}$Fe powers the light curve of stripped-envelope SNe, at maximum light the radiated luminosity is directly related to the amount of $^{56}$Ni synthesised during the explosion \citep{1982ApJ...253..785A}. Fixing the peak bolometric peak luminosity (L$_{p}$), and assuming the rise time (t$_{p}$) being equal to the diffusion time scale ($\tau_{m}$), we can estimate the mass of $^{56}$Ni  using equation 3 of \cite{2016MNRAS.458.2973P}, which is based on the formulation given by \cite{2005A&A...431..423S}. Using L$_{p}$($uBVRI$) = 1.5 x 10$^{42}$ erg sec$^{-1}$, we estimate the $^{56}$Ni mass = 0.07 M$_{\odot}$.

\cite{2016MNRAS.458.2973P} also show that the median NIR contribution to the bolometric peak luminosity is 14 $\%$ and the flux from other wavelengths contribute 10 $\%$. The bolometric luminosity increased to 1.7 x 10$^{42}$ erg sec$^{-1}$  when the contribution from NIR and other wavelengths is also considered which results in $^{56}$Ni mass of 0.08 M$_{\odot}$.

Following \cite{1982ApJ...253..785A}, $\tau_{m}$ determines the width of the bolometric light curve and can be expressed in terms of opacity ($\kappa$), ejecta mass M$_{ej}$, and the photospheric velocity at luminosity peak v$_{ph}$: 
\begin{equation}
\tau_{m} = \sqrt{2}{\bigg(\frac{k}{\beta
c}\bigg)^{\frac{1}{2}}}{\bigg(\frac{M_{ej}}{v_{ph}}\bigg)^{\frac{1}{2}}}
\end{equation}
\noindent
where $\beta=13.8$ is a constant of integration \citep{1982ApJ...253..785A} and {\it c} is the speed of light. Additionally, we assume a constant opacity $k=0.07$ cm$^{2}$ g$^{-1}$ which is justified if electron scattering is the dominant opacity source \citep{1992ApJ...394..599C}. The kinetic energy for spherically symmetric ejecta with uniform density is: 
\begin{equation}
E_{k} = \frac{3}{10} M_{ej} v_{ph}^{2}
\end{equation}
\noindent
In the case of SN 2015as, using $\tau_{m}$ = 17 days and v$_{ph}$ = 7000 km sec$^{-1}$,  we estimate M$_{ej}$ = 2.2 M$_{\odot}$ and E$_{k}$ = 0.656 x 10$^{51}$ erg. \cite{2015ApJ...811..147F} use hydrodynamical models to fit the  bolometric light curves of SNe 2008ax and 2011dh and generate various model combinations of explosion energy, He core mass and $^{56}$Ni mass. A comparison with the models of \cite{2012ApJ...757...31B} and \cite{2015ApJ...811..147F} suggests  an He core mass of 3.3 M$_{\odot}$ in SN 2015as for $^{56}$Ni mass = 0.08 M$_{\odot}$ and E$_{k}$ = 0.656 x 10$^{51}$ erg. The probable ZAMS mass in this scenario is predicted to be between 12 - 15 M$_{\odot}$. 

\begin{figure}
	\begin{center}
		%\vspace{-1.0cm}
		\includegraphics[scale=0.43]{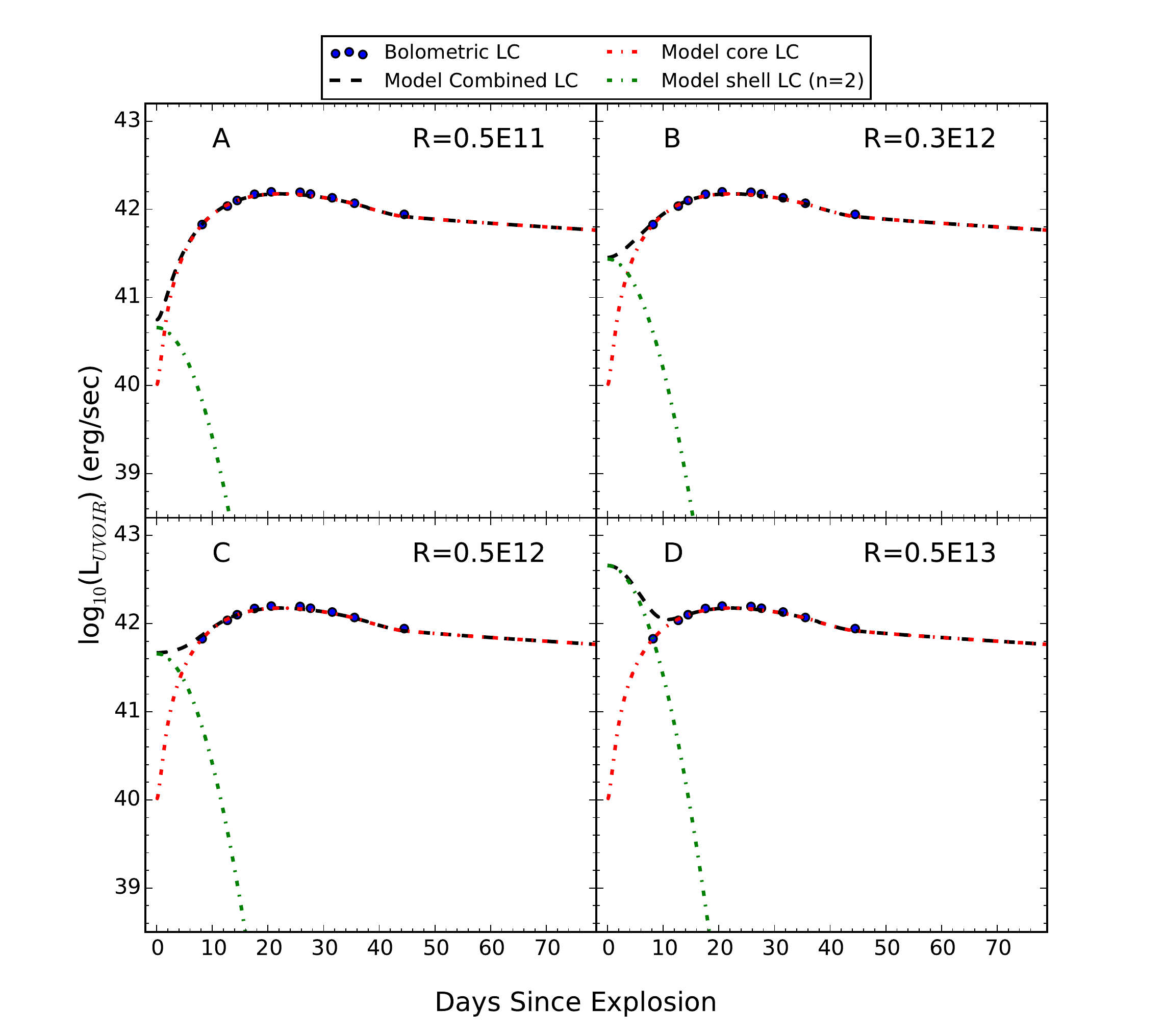}
	\end{center}
	\caption{Bolometric light curves at different values of shell radius. The second and third panel shows that a small variation in radius incorporates the contribution of the shell to the bolometric light curve.}.
	\label{fig:compbol}
\end{figure}

The light curves of SNe IIb are usually modelled by a two-component ejecta configuration, with an extended, low-mass, H-rich outer envelope and a denser and compact He-rich core \citep{2014A&A...571A..77N, 2016A&A...589A..53N}. The  light curve is thus the combination of radiation coming from the shock-heated ejecta and the radioactive decay of $^{56}$Ni to $^{56}$Co. However, in SNe IIb without the early light-curve peak the $^{56}$Ni decay is the dominant source powering the light curve. We assume that SN 2015as is one of these cases. With the assumption that the outer layer retained some H \citep{1989ApJ...340..396A}, $\kappa$ = 0.4 cm$^{2}$ g$^{-1}$ is selected as the Thompson scattering opacity for this layer, while the core is assumed to be composed only by He, and has $\kappa$ = 0.24 cm$^{2}$ g$^{-1}$. The model given by \citet{2014A&A...571A..77N, 2016A&A...589A..53N} solves a set of differential equations through simple numerical integration. As the photon diffusion time scale is much smaller in the outer shell than in the core, the contribution of the two regions to the overall light curve is well separated.  With the two-component model, we are able to extract information on the radius of both the H-rich shell and the He-rich core. 
Using the {\it uBVRI} and full bolometric light curves we find the best-fit model.
In order to do so,  we vary the shell radius on a wide scale. The contribution of the core and shell at different values of shell radius (while keeping other parameters fixed) is shown in Fig. \ref{fig:compbol}. The transition of the shell radius from 0.3 x 10$^{12}$ cm to 0.5 x 10$^{12}$ cm includes the contribution of both the core and the shell to the overall bolometric light curve. Thus, we claim that for a shell radius of 0.5 x 10$^{12}$ cm a best fit model to the full bolometric light curve is obtained (Fig. \ref{fig:compbol} (B)). The best fit values of $^{56}$Ni, M$_{ej}$ (core and shell), E$_{k}$, E$_{Th}$ (Thermal energy), ionisation temperature T and radius of the core and shell,  using the {\it uBVRI} and full bolometric light curve, are given in Table \ref{tab:Nagy}. Using the {\it uBVRI} and full bolometric light curve, our best-fit model yields $^{56}$Ni mass to be 0.07 and 0.08 M$_{\odot}$, respectively, which are consistent with the estimates obtained using Arnett's formulation as discussed above. However, these parameters are uncertain when the light curve is poorly sampled during its cooling phase. We remark that independent values of opacity, ejected mass and kinetic energy cannot be obtained with this method. Only degenerate combinations like M$_{ej}$$\kappa$ and E$_{k}$$\kappa$ can be constrained by observations. Secondly, for the case of stripped-envelope SNe, the values of M$_{ej}$ and E$_{k}$ may change significantly when the tail or the peak of the light curves are fitted. The total kinetic energy E$_{k}$ is the combination of the core and shell.    

Comparing with the results obtained with the Arnett's approximation, we found that the estimated values of E$_{k}$ differ by 0.5 x 10$^{51}$ erg and M$_{ej}$ is well within comparable range while $^{56}$Ni mass also differs by 0.01 M$_{\odot}$.  However, Arnett's formulation works well only when the radius (R) is negligible. For SN 2015as, we estimate a compact radius of 2 x 10$^{11}$ cm ($\sim$ 3 R$_{\odot}$) for the He-rich core and 5 x 10$^{11}$ for the H-rich shell. On the other hand to shock heated ejecta, at the H-recombination front, the thermal energy also contributes in determining the light curve shape \citep{1982ApJ...253..785A, 2014A&A...571A..77N} or else the assumption that the rise time being equal to diffusion time scale, would thus lead to a significant under-estimate of M$_{Ni}$ and E$_{k}$. Also, a minor difference is noticed in the values of M$_{ej}$ as obtained from Arnett's formulation and Nagy $\&$ Vinko models, likely due to the discrepant values of $\kappa$. Arnett assumes that the electron-scattering opacity is the  dominant source of opacity, we therefore vary the opacity between 0.07 to 0.2 cm$^{2}$ g$^{-1}$ and estimate M$_{ej}$ being between 1.1 to 2.2 M$_{\odot}$.  


%%%%%%%%%%%%%%%%%%%%%%%%%%%%%%%%%%%%%%
\section{Spectroscopic Results}
\label{4}
We present the spectral sequence of SN 2015as from $\sim$ 11 days to $\sim$ 230 days after the explosion. The first spectrum was taken on 2015 November 17. For the sake of clarity, line identification and spectral evolution at different phases will be discussed in three separate sections.

\subsection{Pre-Maximum Spectroscopic Features}
\begin{figure*}
	\begin{center}
		%\hspace{-2.55cm}
		\includegraphics[scale=0.50]{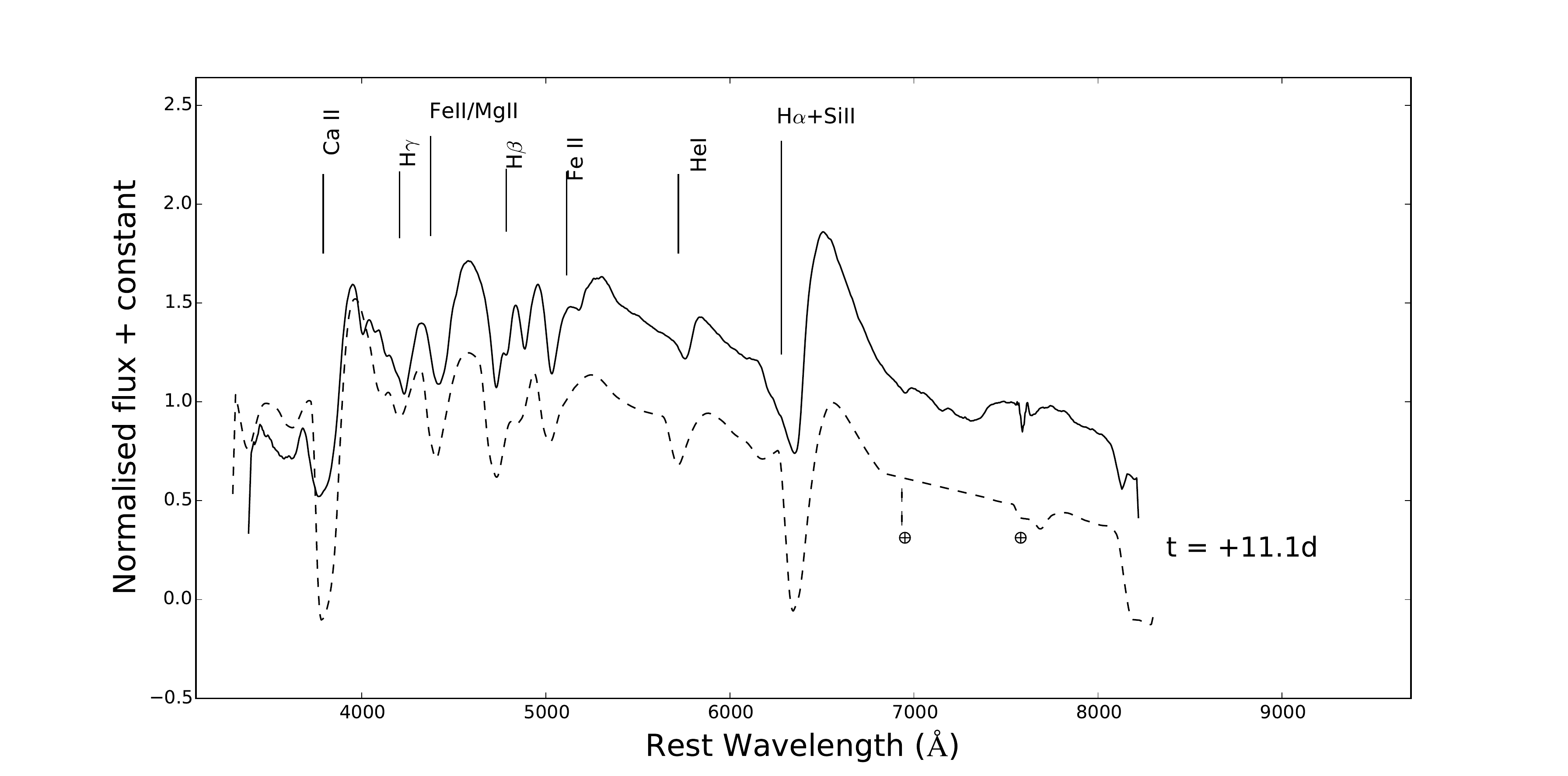}
	\end{center}
	\caption{Early spectrum of SN 2015as (11 days after the explosion). The synthetic spectrum generated using SYN++ modelling is plotted with a dashed line.}
	\label{fig:day1}
\end{figure*}

Fig \ref{fig:day1} shows the pre-maximum spectrum of SN 2015as, obtained  11 days after the explosion. The spectrum has dominant absorption features superimposed on the continuum. The spectrum is dominated by features of Ca II H $\&$ K (3968 and 3933 \AA), P~Cygni absorption profiles of H$\alpha$ (6563 \AA), H$\beta$ (4861 \AA), H$\gamma$ (4341 \AA) and Fe II (4924 and 5018 \AA)/ Mg II lines. The dip at 4882 \AA~ is most likely due to the combination of He I (4921, 5051 \AA) and Fe II (4924 \AA) features. We also identify a weak Fe II absorption line at 5038 \AA. The feature seen at 5760 \AA~is very likely due to He I 5876 \AA.  

The observed spectrum is compared with a synthetic spectrum generated using SYN++ in Fig \ref{fig:day1}. SYN++ is an improved version of the SYNOW code \citep{2011PASP..123..237T}. The basic assumptions of this model are spherical symmetry, local thermodynamic equilibrium for level populations, resonant scattering line formation above a sharp photosphere and homologous expansion of the ejecta ($v$ $\propto$ $r$).  The line formation is treated using the Sobolev approximation \citep{1957SvA.....1..678S,1990ApJ...352..267J}. The optical depth of the strongest line is the free fitting parameter, while the optical depths of other lines of the same ion are determined assuming Boltzmann equilibrium at the excitation temperature T$_{exc}$. However, one should account for approximations due to assuming a  sharp photosphere emitting like a blackbody, and no electron scattering.

The main species required to match the observed spectrum are  marked in Fig \ref{fig:day1}.
The photospheric temperature of the synthetic spectrum at 11 days is 8,000 K. The photospheric velocity obtained from the synthetic spectrum is {\it v$_{ph}$} = 8,500 km sec$^{-1}$. The H lines are detached from the photosphere by a modest amount and with a minimum velocity of {\it v$_{min}$} = 8,800 km sec$^{-1}$. The H$\beta$ and H$\gamma$ features are reproduced well with a minimum velocity {\it v$_{min}$} = 8,800 km sec$^{-1}$ and maximum velocity {\it v$_{max}$} = 15,000 km sec$^{-1}$. The broad H$\alpha$ feature is difficult to fit due to the possible blend with a nearby component. \cite{1995A&AS..110..513B} suggest that the broadness of H$\alpha$ in SN 1993J could be due to a blend of H$\alpha$ and Fe II lines whereas in SN 2000H it could be a blend with C II 6580 \AA~ and Si II 6355 \AA~ line \citep{2006A&A...450..305E, 2009ApJ...703.1624M, 2009PASP..121..689S}. Two components of H with different velocities reproduce the broad H$\alpha$ profile in SN 2011ei \citep{2013ApJ...767...71M}. However, in SN 2015as spectral modelling shows that a two-component H$\alpha$ profile (one being a high velocity component) does not fit well. On the other hand, a combination of H$\alpha$ and Si II 6355 \AA~ reproduces the observed features at 6353 \AA~ (see Fig. \ref{fig:day1}). The feature at 5760 \AA~ is due to He I, as also observed in SN 2008ax \citep{2008MNRAS.389..955P}. 

\begin{figure*}
	\begin{center}
		%\resizebox{20.0cm}{10.0cm}{\includegraphics{spectra2014dt1.pdf}}
		\hspace{-1.0cm}
		\includegraphics[scale=0.45]{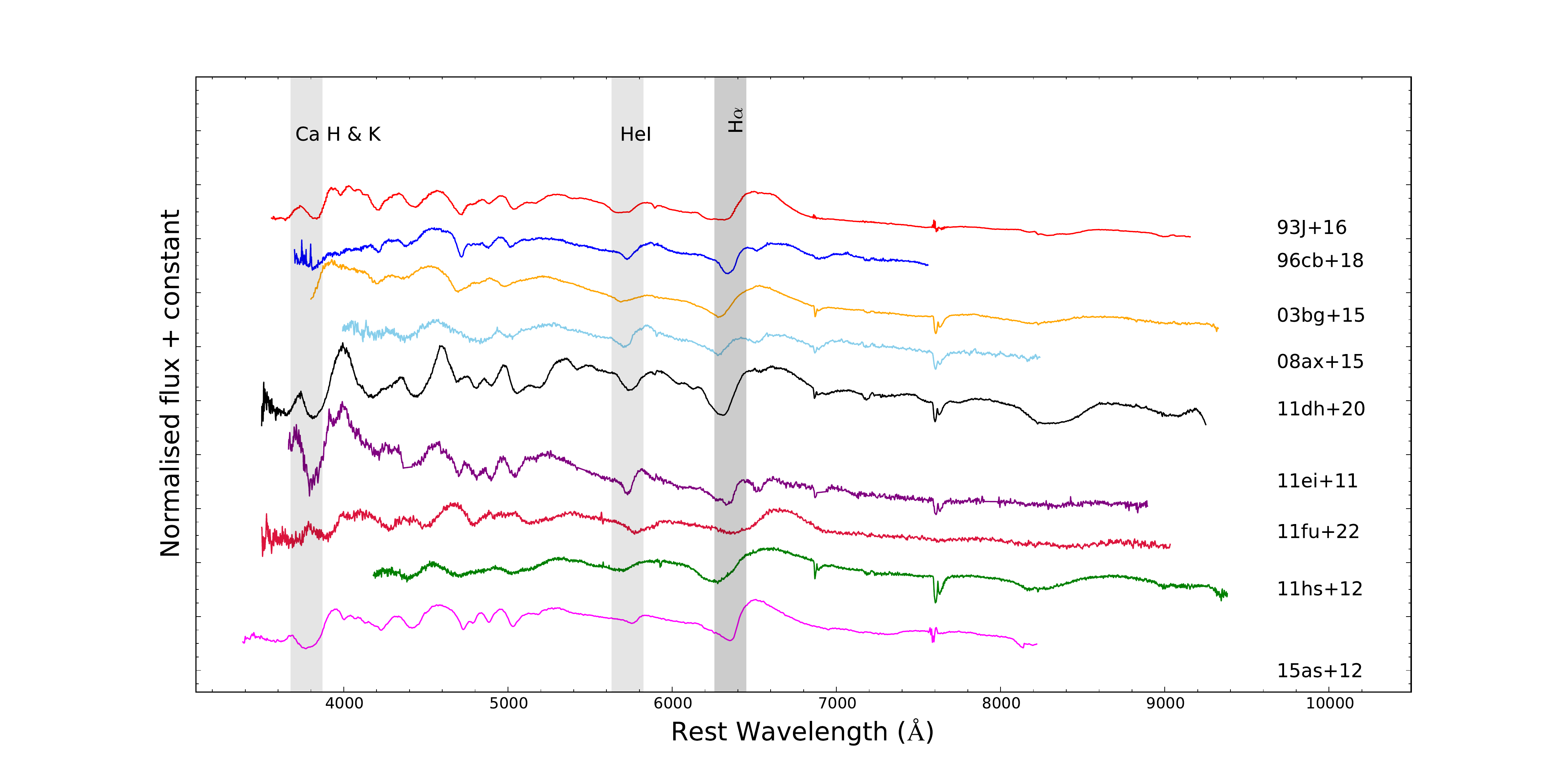}
	\end{center}
	\caption{A comparison of the pre-maximum spectrum of SN 2015as with those of SNe 1993J, 1996cb, 2003bg, 2008ax, 2011dh, 2011ei, 2011fu and 2013df. Even though He I lines are visible, all spectra show strong H$\alpha$ features.}
	\label{fig:comp15-16}
\end{figure*}

A comparison of the early spectrum of SN 2015as with other SNe IIb is shown in Fig \ref{fig:comp15-16}.
The  sample includes SNe 1993J \citep{2000AJ....120.1487M, 1995A&AS..110..513B}, 1996cb \citep{2014AJ....147...99M}, 2003bg \citep{2009ApJ...703.1612H}, 2008ax \citep{2011MNRAS.413.2140T,2008MNRAS.391L...5C,2014AJ....147...99M}, 2011fu \citep{2015MNRAS.454...95M, 2013MNRAS.431..308K}, 2011dh \citep{2013MNRAS.433....2S}, 2011ei \citep{2013ApJ...767...71M}, 2011hs \citep{2014MNRAS.439.1807B} and 2013df \citep{2014MNRAS.445.1647M}. The H$\alpha$ absorption feature is very prominent in all spectra though with different shapes. It is important to note that the He I 5876 \AA~ feature is very weak in SN 2015as as compared to other members of our sample.

\subsection{Transition Phase}
\begin{figure*}
	\begin{center}
		%\resizebox{20.0cm}{10.0cm}{\includegraphics{spectra2014dt1.pdf}}
		\includegraphics[scale=0.45]{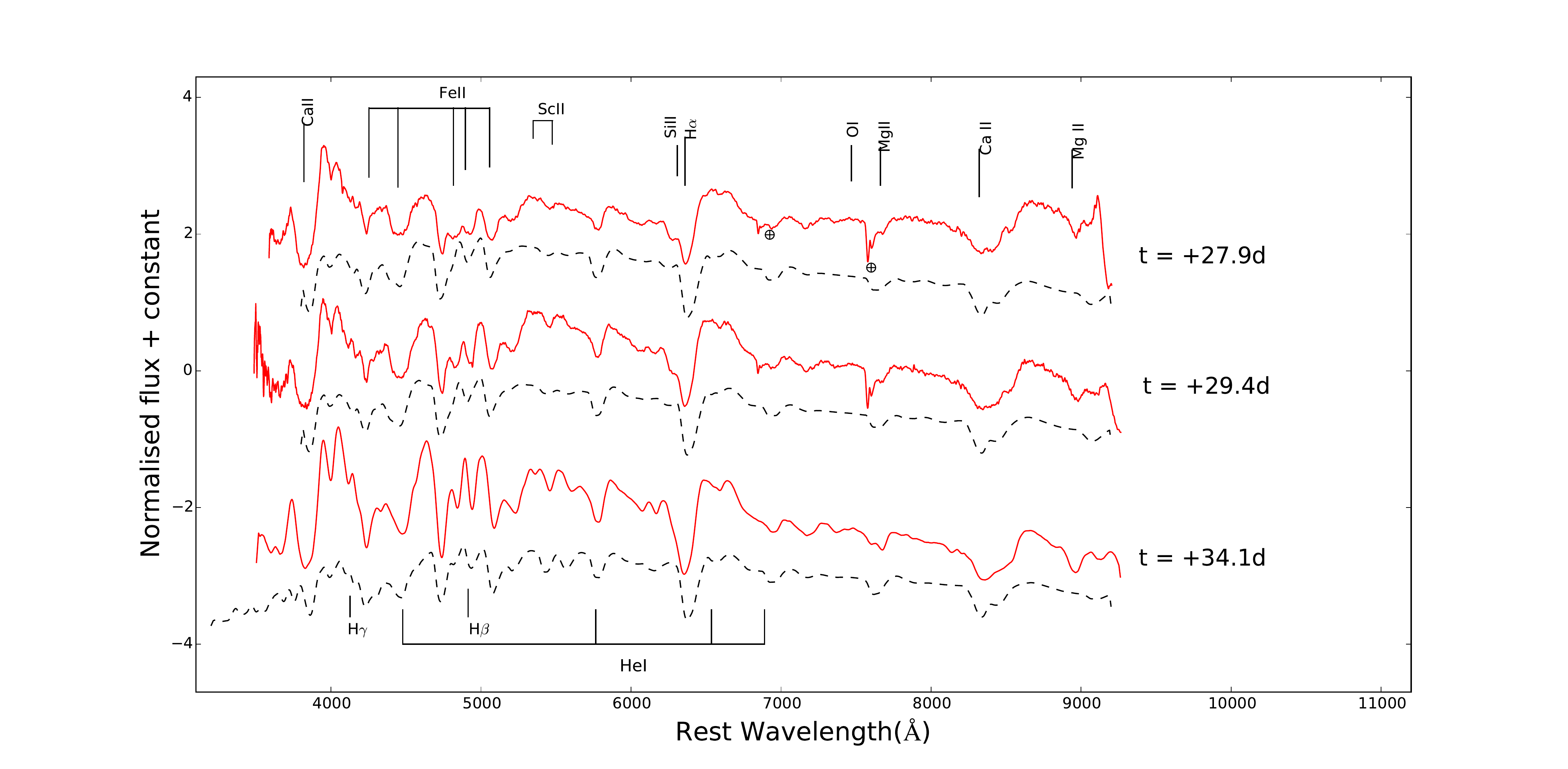}
	\end{center}
	\caption{The 27, 29 and 34 day spectra of SN 2015as are characterised by the appearance of He I features. Phases are calculated from the explosion date. All the spectra show strong H$\alpha$ feature and He I feature with varying strength. The synthetic spectra generated using SYN++ are marked with dashed lines. Spectra are vertically shifted by an arbitrary amount for clarity.}
	\label{fig:day1-2}
\end{figure*}

The second spectrum was obtained 27 days after the explosion, and is shown along with the  29 and 34 day spectra in Fig \ref{fig:day1-2}.
The most prominent lines, including  the He features, are  marked in the figure. A common property of the three spectra is the weakening of the blue continuum. We also note a decrease in the strength of H$\beta$ and H$\gamma$ while He I features now become prominent. In particular, P~Cygni features due to He I lines 4471, 5015, 5876, 6678, 7065 and 7281 \AA~ are now strong. The appearance and the strengthening of He I 6678 \AA~ and 7065 \AA~ about a month after the explosion, indicates that the progenitor of SN 2015as was partially stripped. The  Fe II lines between 4300 and 5000 \AA~ become stronger with time. A double notch in absorption  is seen at 5300 \AA~ probably due to Sc II lines. We also see a O~I 8448 \AA~ line near the left edge of Ca II NIR triplet. The O I line later on blends with the increasing Ca II NIR. The H$\alpha$ absorption line becomes somewhat narrower, and a double trough is seen in 27 and 29 days spectra. The red component is primarily due to the P~Cygni profile of H$\alpha$ line while the origin of the blue component is still not clear. It has been suggested that the blue component is also a part of H$\alpha$ line, e.g. due to the presence of non-spherical density distribution of H \citep{1993Natur.364..600S} or the existence of a second high velocity H$\alpha$ layer \citep{1995MNRAS.277.1115Z, 2002ApJ...566.1005B}. The spectral modelling reveals again that a combination of H$\alpha$ and Si II 6355 \AA~ reproduces the observed features at 6353 \AA. 

A close look at the 27, 29 and 34 day spectra reveals that the H profile becomes sharper. The synthetic spectra using SYN++ with $v_{max}$ decreasing from 12,000 km sec$^{-1}$ to 7,000 km sec$^{-1}$, and the photospheric temperatures fading from 8,000 to 6,000 K fit the observed spectra well. The He I 5876 \AA~ feature is well fitted, and the blueshift of its absorption profile indicates velocities decreasing from 8,000 to 7,000 km sec$^{-1}$. The O I, Ca II and Mg II lines become prominent, with velocities 7,500, 9,000 and 8,000 km sec$^{-1}$, respectively.

\begin{figure*}
	\begin{center}
		%\resizebox{20.0cm}{10.0cm}{\includegraphics{spectra2014dt1.pdf}}
		\includegraphics[scale=0.50]{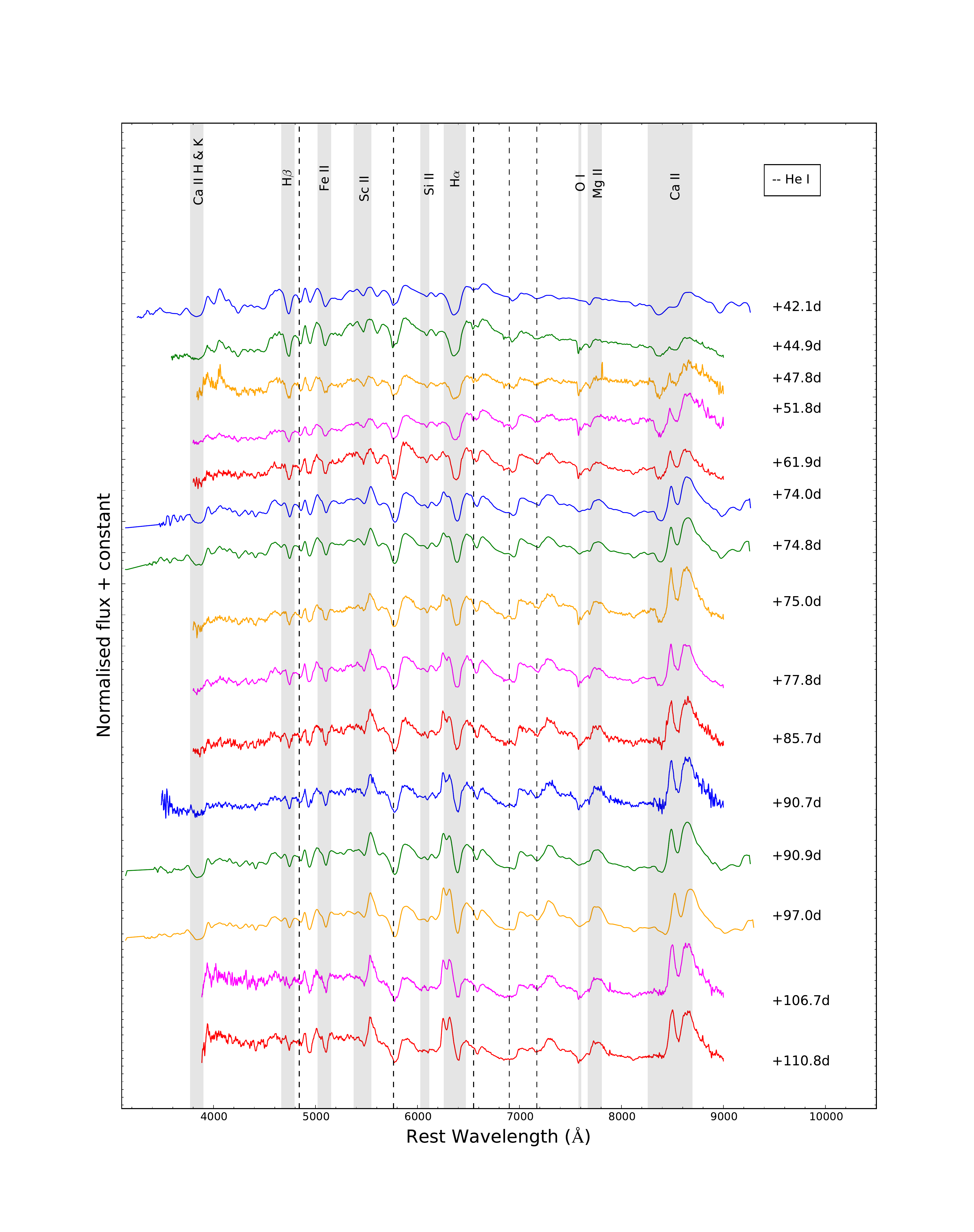}
	\end{center}
	\caption{Spectral evolution of SN 2015as from 42 to 110 days after explosion. The important lines are marked by shaded regions. Dotted lines represents the prominent He I features at different stages of the SN evolution.}
	\label{fig:midspectra}
\end{figure*}

\begin{figure}
	\begin{center}
		%\hspace{-4.0cm}
		%\resizebox{20.0cm}{10.0cm}{\includegraphics{spectra2014dt1.pdf}}
	%	\hspace{-1.5cm}
		\includegraphics[scale=0.21]{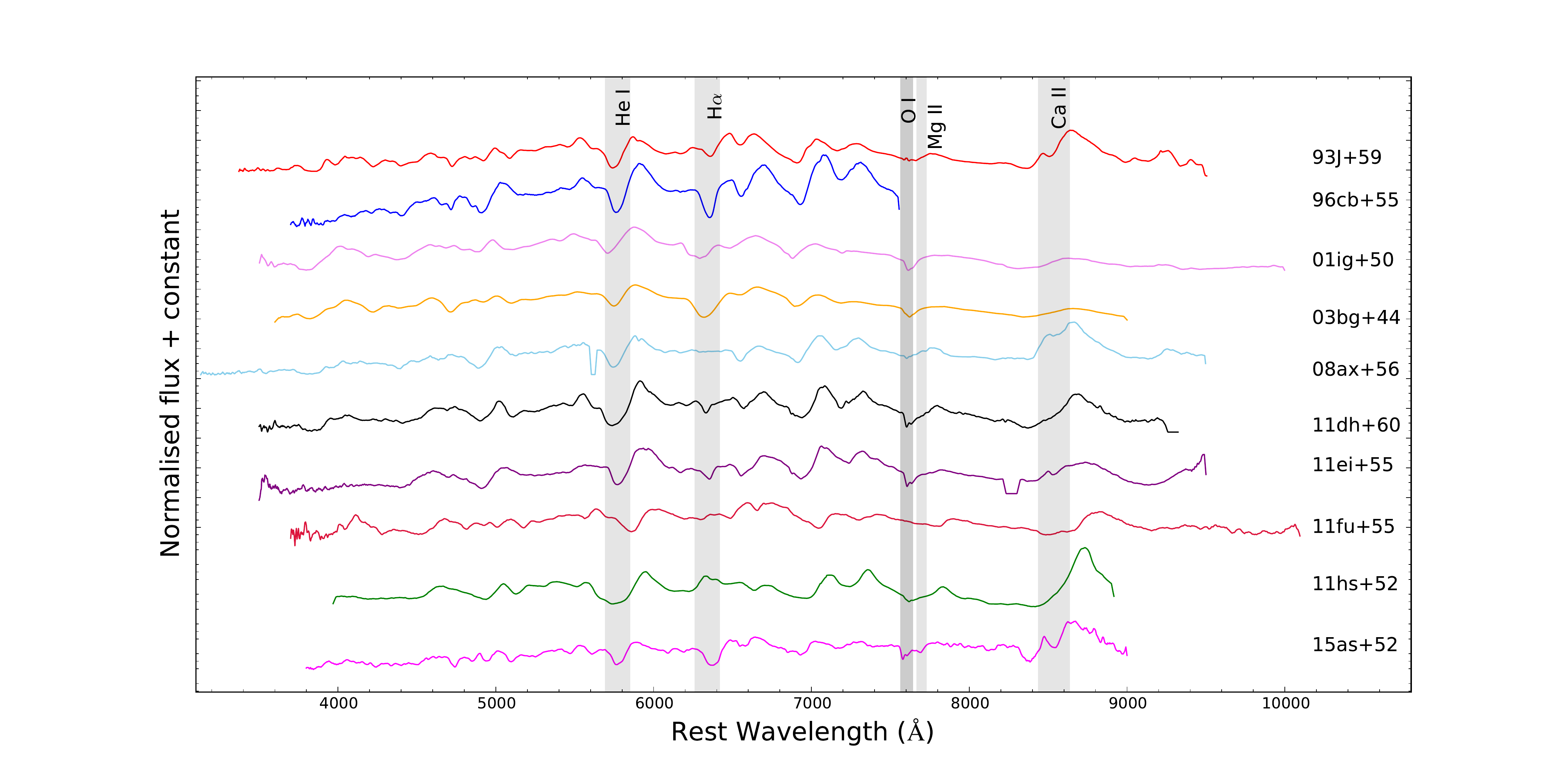}
	\end{center}
	\caption{A comparison of SN 2015as and  other SNe IIb at intermediate epochs ($\sim$ 45-60 days) of their evolutions. The spectra show weakening of H lines along with dominant Ca II NIR and He I features. The important lines are marked by shaded regions.}
	\label{fig:comp50-60}
\end{figure}

Subsequent spectra, in the transition from the photospheric to the nebular phase, show a weakening of the Ca II H $\&$ K feature (see Fig \ref{fig:midspectra}).
The absorption of H$\alpha$ decreases in strength with time and, at $\sim$ 75 days, it marks the metamorphosis of the SN spectrum from Type II to Type Ib. Lines of He I 4471, 5015, 5876, 6678, 7065 and 7281 \AA~ are now well observed. The presence of H$\alpha$ and H$\beta$ absorption features until $\sim$ 75 days indicates that the progenitor was still relatively H-rich at the time of the explosion. A residual H$\alpha$ emission feature is seen between 75 to 110 days post explosion. The O I line at 7774 \AA~ is detected, in addition to the absorption due to Mg II. The He I 5876 \AA~ wing is becoming prominent and the He I absorption profile shows a notch due to the Na ID feature. This phase is characterised by the appearance of the Ca II NIR emission feature, which gradually becomes more prominent from the 50 to 110 days spectra. The spectra now resemble those of a Type Ib event, except for the residual presence of the H$\alpha$ feature. The forbidden line of [O I] 5577 \AA~, and O I at 7774 \AA~ become more prominent with time. Fig \ref{fig:comp50-60} shows that the spectrum of SN 2015as at this phase match well those of SNe 1993J, 2008ax and 2011dh. 
The H$\alpha$ profile at this stage is similar to SNe 2003bg and 1996cb, suggesting a similar progenitor configuration. The Ca II NIR emission of SN 2015as matches those observed in the spectra of SNe 1993J and 2008ax, although the peaks are more pronounced in SN~2015as.

\subsection{Nebular Phase}
\begin{figure}
	\begin{center}
		\includegraphics[scale=0.21]{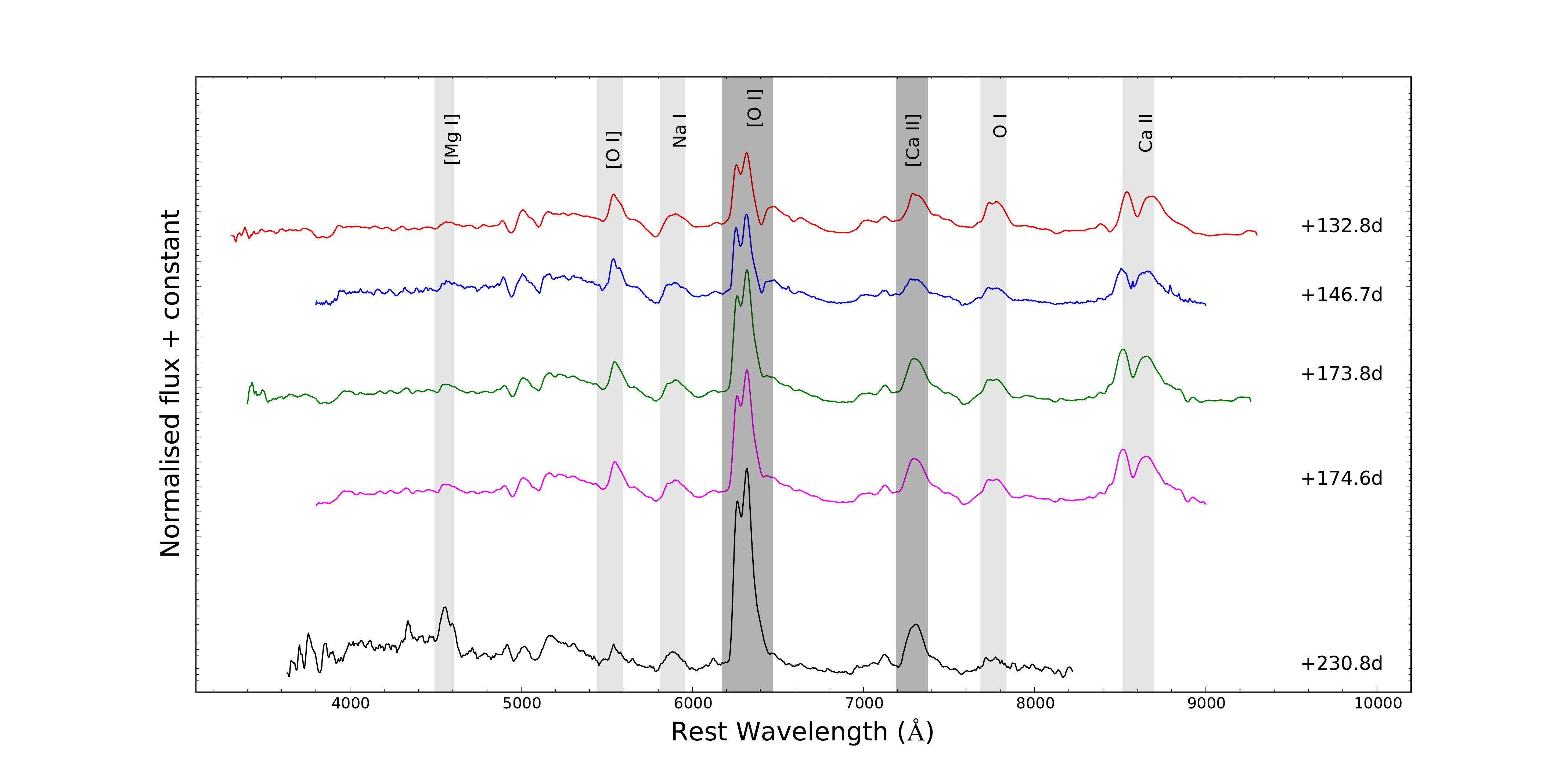} \\
	\end{center}
	\caption{Nebular phase spectral evolution of SN 2015as from 132 to 230 days post explosion. Prominent lines such as the [O I] and [Ca II] doublets, along with [Mg I] singlet, are marked by shaded regions.}
	\label{fig:day171819}
\end{figure}

\begin{figure}
	\begin{center}
		%\hspace{-3.17cm}
		\includegraphics[scale=0.21]{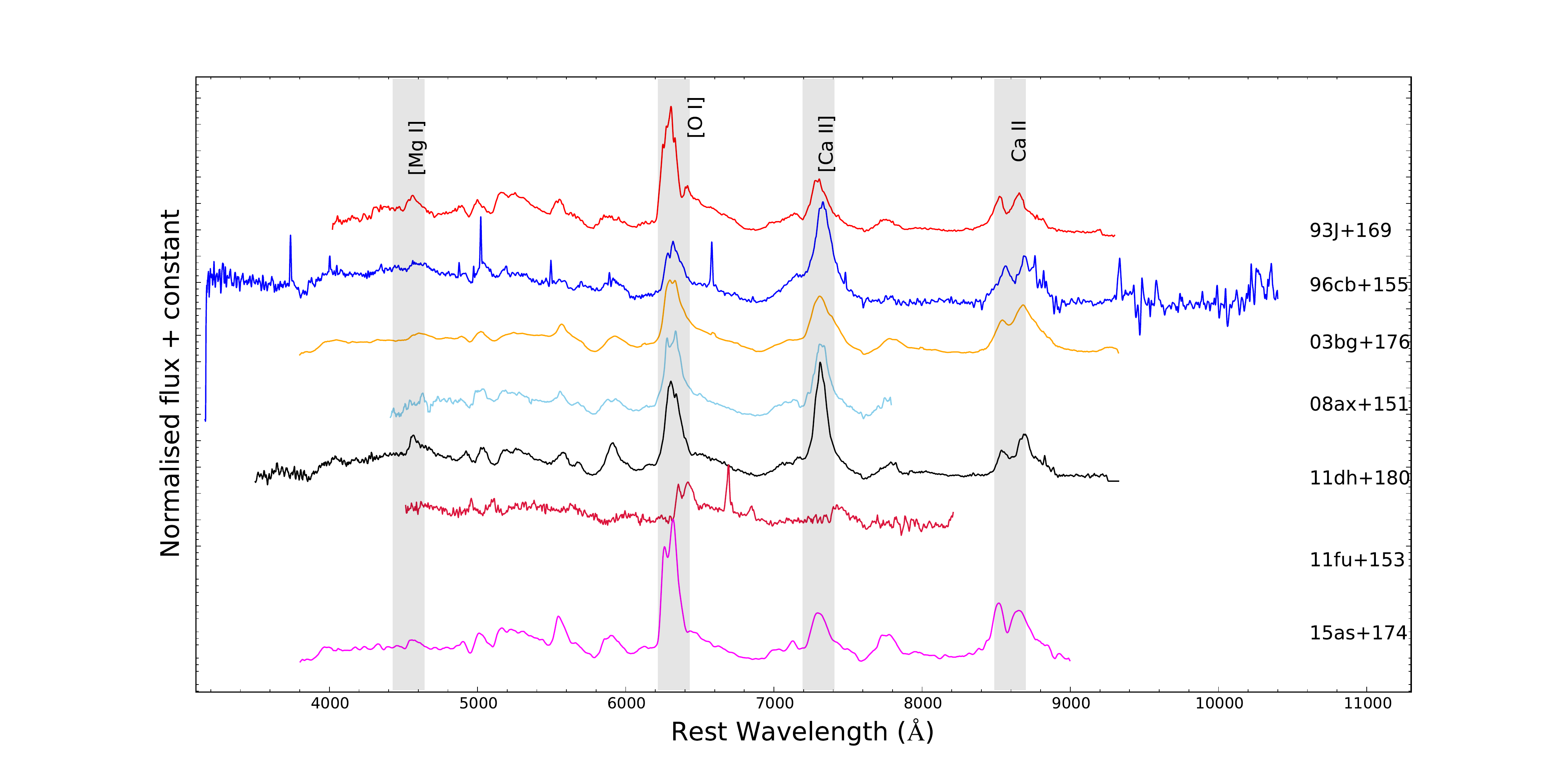} \\
	\end{center}
	\caption{Comparison of the nebular phase spectrum of SN 2015as with other SNe IIb. Prominent doublet of [O I] is marked.}
	\label{fig:comp150}
\end{figure}

The nebular phase spectral evolution from 132 to 230 days after explosion is shown in the Fig \ref{fig:day171819}. 
The nebular spectra are dominated by the strong [O I] doublet at 6300, 6364 \AA. The 230 days spectrum also shows a prominent [Mg I] 4571 \AA~ line, which has a similar origin as the [O I] doublet. A broad emission likely due to O I 7774 \AA~ is also seen up to 230 days after explosion. The nebular phase spectra of SNe IIb are compared in Fig \ref{fig:comp150}. 
We note that  H$\alpha$ has faded in all SN spectra which are dominated by emission lines of [Mg I] 4571 \AA, [O I] 6300, 6364 \AA, [Ca II] 7291, 7324 \AA, OI 7774 \AA, blends of [Fe II] lines near $\sim$ 5000 \AA~ and the Ca II NIR triplet. The weakening of [O I] 5577 \AA~ line can be noticed. Although nebular features are similar for all SNe , differences exist with respect to the relative strengths of emission lines, and their shapes. The Ca II NIR triplet in SN 2015as is very similar to that of SN 1993J, although the overall spectrum closely matches those of SNe 2008ax and 2011dh. The Ca II NIR feature in SN 2015as has a different emission geometry when compared with SN 2003bg and SN 2011dh (Fig \ref{fig:comp150}). Although the He I lines at 6678, 7065 \AA~ are now relatively weaker, the He I 5876 \AA~ is still pronounced in SN 2015as and SN 2011dh, most likely due to an increasing contribution of Na ID. We also estimate the [Mg I]/[O I] flux ratio on the last spectrum of SN 2015as, and found the value to be 0.23. \cite{2013MNRAS.433....2S} found that at $\sim$ 280 days post explosion, the ratio is $\sim$ 0.2 in SN 2011dh, $\sim$ 0.15 in SN 1993J and 0.36 in SN 2001ig, which has the strongest [Mg I] line \citep{2009PASP..121..689S}. The emission features of [O I] 6300, 6363 \AA~, [Ca II] 7291, 7324 \AA~ doublets and [Mg I] 4571 \AA~ singlet are discussed in more detail below. 

The H$\alpha$ absorption is detected up to 146 days in SN 2015as, and completely disappears thereafter. We notice that  H$\alpha$ is seen in the nebular phases (beyond 150 days) in SN 1993J, which is plausibly a signature of circumstellar interaction \citep{2000AJ....120.1499M, 2010MNRAS.409.1441M}, or due to mixing and clumping of H and He ionised by the radioactive decay \citep{2010MNRAS.409.1441M} or even blending of H$\alpha$ and [N II] 6548, 6584 \AA. In the case of SN 2015as, no nebular H$\alpha$ emission is observed after 150 days, implying that circumstellar interaction is not significant. 

\begin{figure*}
	\begin{center}
		\hspace{-1.0cm}
		\includegraphics[scale=0.42]{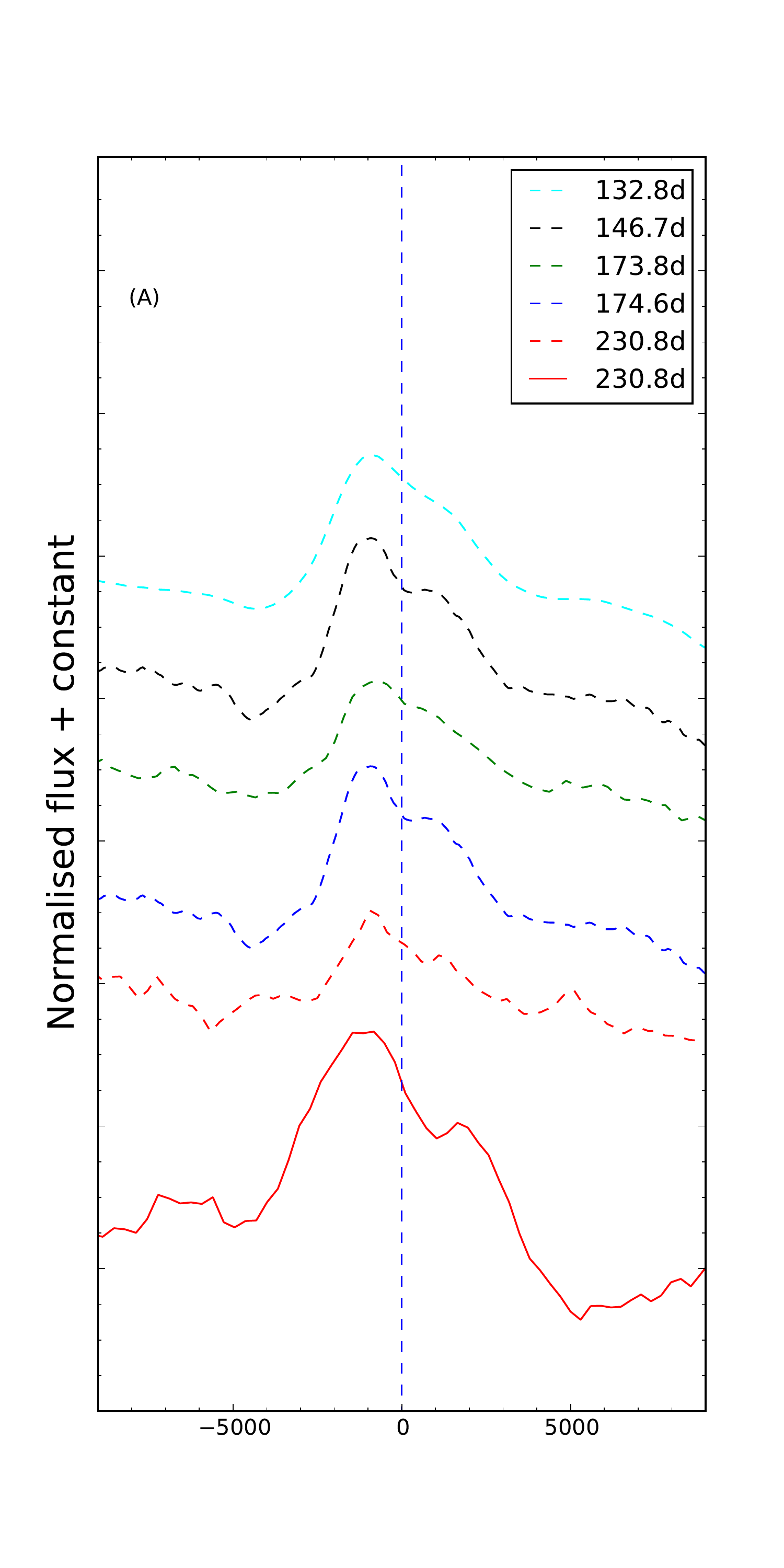} 
		\hspace{-0.8cm}
		\includegraphics[scale=0.42]{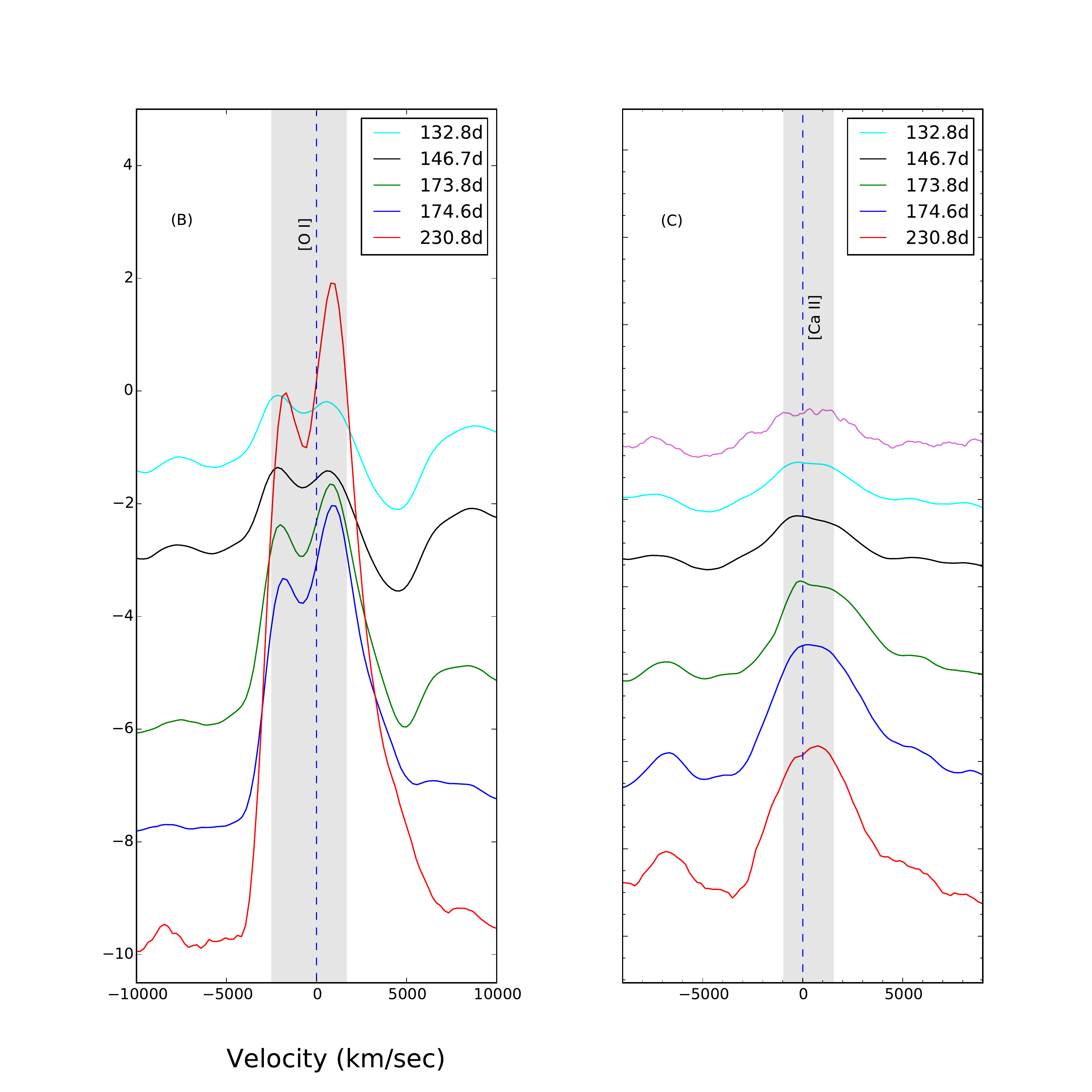} \\
	\end{center}
	\caption{Line profiles of [Mg I] 4571 \AA/ [O I] 5577 \AA, [O I] 6300, 6364 \AA~ and [Ca II] 7291, 7324 \AA~ during the nebular phase. The first five dashed lines in panel (A) represent the evolution of the [O I] 5577 \AA~ line, while the last one (solid red line) is for [Mg I] 4571 \AA.}
	\label{fig:comb_vel}
\end{figure*}

{\underline{\bf Magnesium Emission at Late Phases:}} \cite{2006ApJ...640..854M}, through their hydrodynamic explosion models, suggest that [Mg I] and [O I] have similar spatial evolution within the SN ejecta. A direct comparison of the [Mg I] and [O I] 6300, 6364 \AA~profiles is problematic as the [O I] feature is a doublet. However, we compare the evolution of the [O I] 5557 \AA~ with [Mg I] 4571 \AA. in the case of SN 2015as, the [Mg I] feature is not as prominent as [O I] in the early nebular spectra. This may be due to different geometry originating in the stratification of progenitor star and the hydrodynamics of the explosion. The [Mg I] 4571 \AA~ feature at day 230 shows an asymmetric profile similar to that of [O I]. In Fig \ref{fig:comb_vel}(A), we present the evolution of [O I] 5577 \AA~ line profile and compare it with that of [Mg I] at 230 days. The two lines have very similar profiles in the late nebular phase. This indicates that the line profiles are governed by the ejecta geometry, and [O I] and [Mg I] have a similar distribution in the SN ejecta.

{\underline{\bf Oxygen Emission at Late Phases:}} The [O I] 6300, 6364 \AA~ usually have a double-peaked structure. Fig \ref{fig:comb_vel}(B) shows the evolution of the double peaked [O I] line from 132 to 230 days after explosion. The [O I] line is becoming prominent and more symmetric with time. The wavelength corresponding to zero velocity is that of the 6300 \AA~ line. The two peaks of the [O I] feature in the SN 2015as spectra are separated by 62 \AA~, and this matches the expectation for the two lines of the [O I] doublet. The absence of [O I] 5577 \AA~ marks the onset of optically thin regime \citep{2009MNRAS.397..677T}, predicted with an intensity ratio for the two lines of the doublet of 3:1 \citep{2001ApJ...559.1047M,2009MNRAS.397..677T}. The observed  profile gives information on the geometry of the emitting region \citep{2009MNRAS.397..677T,2013ApJ...775L..43T}. in the case of SN 2015as, the ratio of [O I] 6300, 6364 \AA~ to [O I] 5577 \AA~ line fluxes is about ~7, hence, the contribution of [O I] 5577\AA~ flux is actually negligible. SN 2015as has one peak close to the rest velocity while the other one is blueshifted by $\sim$ -1700 km sec$^{-1}$.  The possible explanations for the observed blueshift are low mode convective instabilities \citep{2004PhRvL..92a1103S,2006A&A...453..661K} or the suppression of the redshifted part of the emission spectrum caused by dust formation as the ejecta cool \citep{2000AJ....120.1499M,2004A&A...426..963E}. A straightforward geometric explanation could be either asymmetric explosions with the emitting oxygen located in a torus or in a disc perpendicular to the line of sight \citep{2001ApJ...559.1047M,2006ApJ...645.1331M}, or a blob of oxygen moving perpendicularly to the line of sight. For different SNe, multiple explanations were proposed for the origin of double peaked profiles. The [O I] line profile, in most cases is the result of clumpy ejecta with a sawtooth profile  \citep{2000AJ....120.1499M}, which is supported by the explosion scenario of SN 1987A \citep{1993ApJ...419..824L} where the O emission originates from clumps of newly synthesised material, while the Ca emission mainly originated from pre-existing, uniformly distributed material. \cite{1989ApJ...345L..43F} suggested that these enhancements come from Rayleigh-Taylor fingers of high speed material or changes in local density contrasts. Asphericity in the explosion would also result in asymmetric peaks \citep{2010ApJ...709.1343M}. An alternative explanation for the double peaked structure of O lines could be a high-velocity (12,000 km sec$^{-1}$) H$\alpha$ absorption component, which causes a split in the [O I] peak  \citep{2010MNRAS.409.1441M}. Since there is no late-time H$\alpha$ emission in SN 2015as, the contribution of H to the [O I] split can be ruled out. The asymmetric profile of SN 2015as is very similar to that observed for SN 2008ax, that was attributed to a thin O torus viewed from equatorial direction in addition to a spherically symmetric mass of O, or an aspheric distribution of $^{56}$Ni exciting the O layers \citep{2009MNRAS.397..677T}.
 
{\underline{\bf Calcium Emission at Late Phases:}} As shown in Fig \ref{fig:comp150}, the [Ca II] 7291, 7324 \AA~ emission in SN 2015as matches that of SN 1993J. The [Ca II] emission line in SN 2015as has a round peak unlike SN 2008ax and SN 2011dh which exhibit sharp peak. Fig \ref{fig:comb_vel}(C) shows the [Ca II] evolution from 132 to 230 days after the explosion, with the feature increasing in strength with time. The pre-existing envelope excites the Ca mixed in the atmosphere to reach the temperature required to produce the [Ca II] emission, as H and He do not radiate efficiently \citep{1993ApJ...419..824L, 2000AJ....120.1487M}. Calcium clumps are formed during explosion but they do not contribute significantly to the [Ca II] emission. They intercept $\gamma$-ray radiation, however, as the mass fraction is comparatively less, the amount of radioactive luminosity and temperature achieved is not sufficient for the emission.

\subsection{Estimates of the O mass and [Ca II] / [O I] ratio}

The line strengths in the nebular phase spectra provide key information on the progenitor mass. \cite{1986ApJ...310L..35U} derived a relation, showing the minimum mass of O that can be estimated in the high density  (N$_{e}$ $\geq$ 10$^{6}$ cm$^{-3}$) limit. This is given by
\begin{equation}
M_{O} = 10^{8} \times D^{2} \times F([O\,I]) \times exp^{(2.28/T_{4})},
\end{equation} 

\noindent
where $M_{O}$ is the mass of the neutral O in M$_{\odot}$ units, $D$ is distance to the galaxy in Mpc, $F([O I])$ is the total flux of the [O I] 6300,6364 \AA~ feature in erg sec$^{-1}$ cm$^{-2}$, and $T_{4}$ is the temperature of the O-emitting region in units of 10$^{4}$ K. Ideally, the ratio of [O I] 5577 \AA~to the [O I] 6300, 6364 \AA~ feature should be considered. However, the [O I] 5577 \AA~ line is very faint, and the limit of the flux ratio can be assumed to be $\leq$ 0.1. In this limit two conditions exist, either high density (N$_{e}$ $\geq$ 10$^{6}$ cm$^{-3}$) and low temperature  ($T_{4}$ $\leq$ 0.4 K), and low density (N$_{e}$ $\leq$ 10$^{6}$ cm$^{-3}$) and high temperature  (T$_{4}$ = 1.0 K) \citep{2007ApJ...666.1069M}. The O-emitting region is found at high density and low temperature  \citep{1989AJ.....98..577S,1991ApJ...372..531L,2004A&A...426..963E}. Using the observed flux of 4.1 x 10$^{-14}$ erg sec$^{-1}$ cm$^{-2}$ of the [O I] 6300, 6364 \AA~ doublet from the 2016 April 27 spectrum, and adopting T$_{4}$ = 0.4 K, we estimate $M_{O}$ = 0.45 M$_{\odot}$. The [O I] 7774 \AA~ line is also seen in the nebular spectrum, which is mainly due to the recombination of ionised O  \citep{1986ApJ...302L..59B}. This also indicates the presence of O in ionised form, thus $M_{O}$ = 0.45 M$_{\odot}$ can be considered as the lower limit of the O mass ejected during explosion. For the sample of stripped-envelope SNe considered by \citet{2004A&A...426..963E}, the O mass usually ranges from about 0.2 to 1.4 M$_{\odot}$. The estimated O mass is 0.5 M$_{\odot}$ for SN  1993J \citep{1996ApJ...456..811H}, 0.1 - 0.3 M$_{\odot}$ for SN 1996N \citep{1998A&A...337..207S}, 0.2 M$_{\odot}$ for SN 2007Y \citep{2009ApJ...696..713S}, 1.3 M$_{\odot}$ for SN 2003bg \citep{2009ApJ...703.1624M} and  0.22 M$_{\odot}$ for SN 2011dh \citep{2013MNRAS.433....2S}. The O mass for SN 2015as is hence similar to that of SN 1993J.

The O layer formed during the hydrostatic burning phase is responsible for the [O I] emission. The ejected mass of O is directly related to the main sequence progenitor mass. \cite{1996ApJ...460..408T} made explosive nucleosynthesis calculations and predicted major nucleosynthesis yields for the progenitor mass of 13-25 M$_{\odot}$. For progenitor masses of 13, 15, 20 and 25 M$_{\odot}$, \cite{1996ApJ...460..408T} showed that the corresponding O masses would be 0.22, 0.43, 1.48 and 3.0 M$_{\odot}$, respectively. \cite{1996ApJ...460..408T} also estimated He core masses of 3.3, 4 and 8 M$_{\odot}$, corresponding to progenitors of 13, 15 and 25 M$_{\odot}$, respectively. Adopting the O mass estimate for SN 2015as to be 0.45 M$_{\odot}$, the corresponding progenitor mass would be a $\sim$ 15 M$_{\odot}$ with a He core mass $\sim$ 4 M$_{\odot}$. The ratio of the  [Ca II] 7291-7324 / [O I] 6300-6364 fluxes serves as an  indicator for estimating the main sequence mass of the progenitor. The O mass primarily depends on the mass of the progenitor, while Ca is mostly independent of it, so a small ratio of this flux would point to a high progenitor mass \citep{2006NuPhA.777..424N}. Indeed, \cite{1989ApJ...343..323F} have shown that the variations in the composition of the He core mass lead to substantial differences in the spectra, especially in the [Ca II] / [O I] emission-line ratio. The He core masses and thus the progenitor mass can be estimated from such [Ca II] / [O I] ratio. For SN 2015as, the ratio of [Ca II] / [O I] at 174 to 230 days is found to be nearly constant, around 0.44. The value is lower than that measured for SN 1993J ($\sim$ 0.5) with progenitor mass 14-16 M$_{\odot}$, SN 1996cb ($\sim$ 2), SN 2003bg ($\sim$ 0.5) with progenitor mass 22-25 M$_{\odot}$, SN 2008ax ($\sim$ 0.9) with progenitor mass 18.6 M$_{\odot}$, SN 2011hs ($\sim$ 1.2) with progenitor mass 11-15 M$_{\odot}$ and SN 2011dh ($\sim$ 0.8) with progenitor mass 13-18 M$_{\odot}$ at comparable epochs \citep{2011MNRAS.413.2140T, 2001astro.ph..6404D, 2009ApJ...703.1624M, 2011ApJ...739...41C, 2014MNRAS.439.1807B, 2013MNRAS.433....2S}. \cite{2015A&A...579A..95K} compared this ratio for a group of stripped-envelope SNe, and found that it is not sensitive to density and temperature changes, while depends extensively on the progenitor mass. They also found that this ratio remains nearly constant during the nebular phases, and never exceeds the value of $\sim$ 0.7 for Type II SNe, while a considerable diversity exists for Type Ib/Ic events \citep{1989ApJ...343..323F, 2017arXiv170206702J}. The estimated value is similar to that measured for SN 1993J and SN 2003bg. The low flux ratio is indicative of a lower abundance of [Ca II] in the pre-existing envelope. Moreover, the [Ca II] doublet sometime gets blended with [O II] (as in SN 1995N; see \citealt{2002ApJ...572..350F}) and for such cases the ratio of [Ca II]/[O I] may not be a good proxy for core mass estimation. The probable progenitor scenario for SN 2015as is either a 14-15 M$_{\odot}$ star in a binary association like in the case of SN 1993J or a Wolf-Rayet star of 20-25 M$_{\odot}$ mass like SN 2003bg.

\subsection{Velocity Evolution}
\begin{figure}
	\begin{center}
		\includegraphics[scale=0.45]{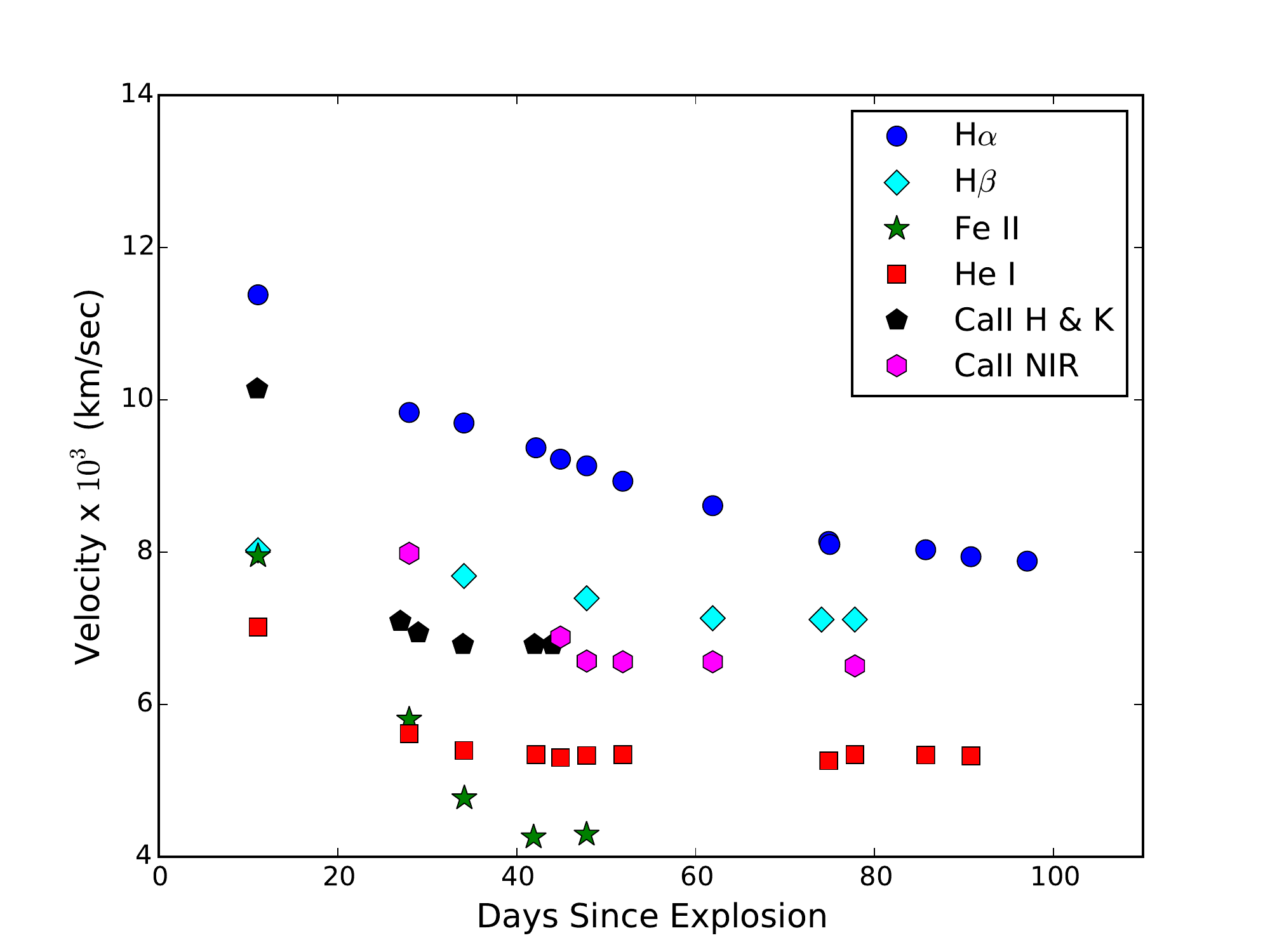} \\
	\end{center}
	\caption{The evolution of the expansion velocity of the SN~2015as ejecta measured from different lines.}
	\label{fig:vel}
\end{figure}

The expansion velocity of the ejecta is usually measured from the absorption minima of the P~Cygni profile by fitting a gaussian. We choose the relatively isolated lines of H$\alpha$, H$\beta$, He I 5876 \AA, Fe II 5169 \AA, Ca H $\&$ K and the Ca II NIR triplet. We choose 3934 \AA~as the rest wavelength for Ca II doublet and 8498 \AA~ for the Ca II NIR triplet. The evolution of the line velocities is shown in Fig. \ref{fig:vel}. The H$\alpha$ line velocity is about 11,300 km sec$^{-1}$ at 11 days after explosion, and fades to $\sim$8,000 km sec$^{-1}$ at 80 days after explosion, becoming almost constant thereafter. However, the late-time velocity estimates are affected by a large uncertainty, as the blue wing of the absorption profile of H$\alpha$ is contaminated by the [O I] emission feature. The velocity of Ca II H $\&$ K at day 11 is 10,100 km sec$^{-1}$, but has a faster decline until $\sim$30 days, to become almost constant later on. The velocity trend of He I is similar to that of H$\alpha$ but with comparatively lower velocities. The measured velocity of He I at $\sim$ 11 days post explosion is 7,000 km sec$^{-1}$, drops to 5,500 km sec$^{-1}$ on day 30 and then remains almost constant. The Ca II NIR feature also has a similar trend starting with a velocity of 8,000 km sec$^{-1}$, decreasing fast and becoming constant at 6,500 km sec$^{-1}$. H$\beta$ follows a gradual decline with an initial velocity of 8,200 km sec$^{-1}$ and becomes constant at $\sim$ 7,000 km sec$^{-1}$ around 77 days post explosion.  Fe II lines are considered a good tracer for the photospheric velocity. The Fe II line in SN 2015as has a velocity of 7,900 km sec$^{-1}$, and declines to 4,200 km sec$^{-1}$ at 45 days past explosion. \cite{2009ApJ...703.1624M}  performed detailed spectral modelling of SN 2003bg in both photospheric and nebular phases revealing the existence of the stratified layers of weaker Balmer lines at 10,000 km sec$^{-1}$ followed by He I lines at 7,000 km sec$^{-1}$. Similarly, it appears that SN 2015as has a three-layer velocity stratification - an outer layer of H moving with high velocity ($\sim$ 8,000 km sec$^{-1}$); an intermediate Ca-rich layer (7000 km sec$^{-1}$) and a final dense iron core with 4,000 km sec$^{-1}$. This scenario can be claimed both from the minima of the absorption profiles that are evolving with time and also from the velocities obtained from the SYN++ modelling. SN 2011dh also showed a similar velocity profile \citep{2013MNRAS.433....2S}, although the velocities in SN 2011dh were slightly higher than in SN 2015as. It is also important to note that not only stratification but optical depth also plays a vital role in this division of velocities.

\begin{figure}
	\begin{center}
		\includegraphics[scale=0.40]{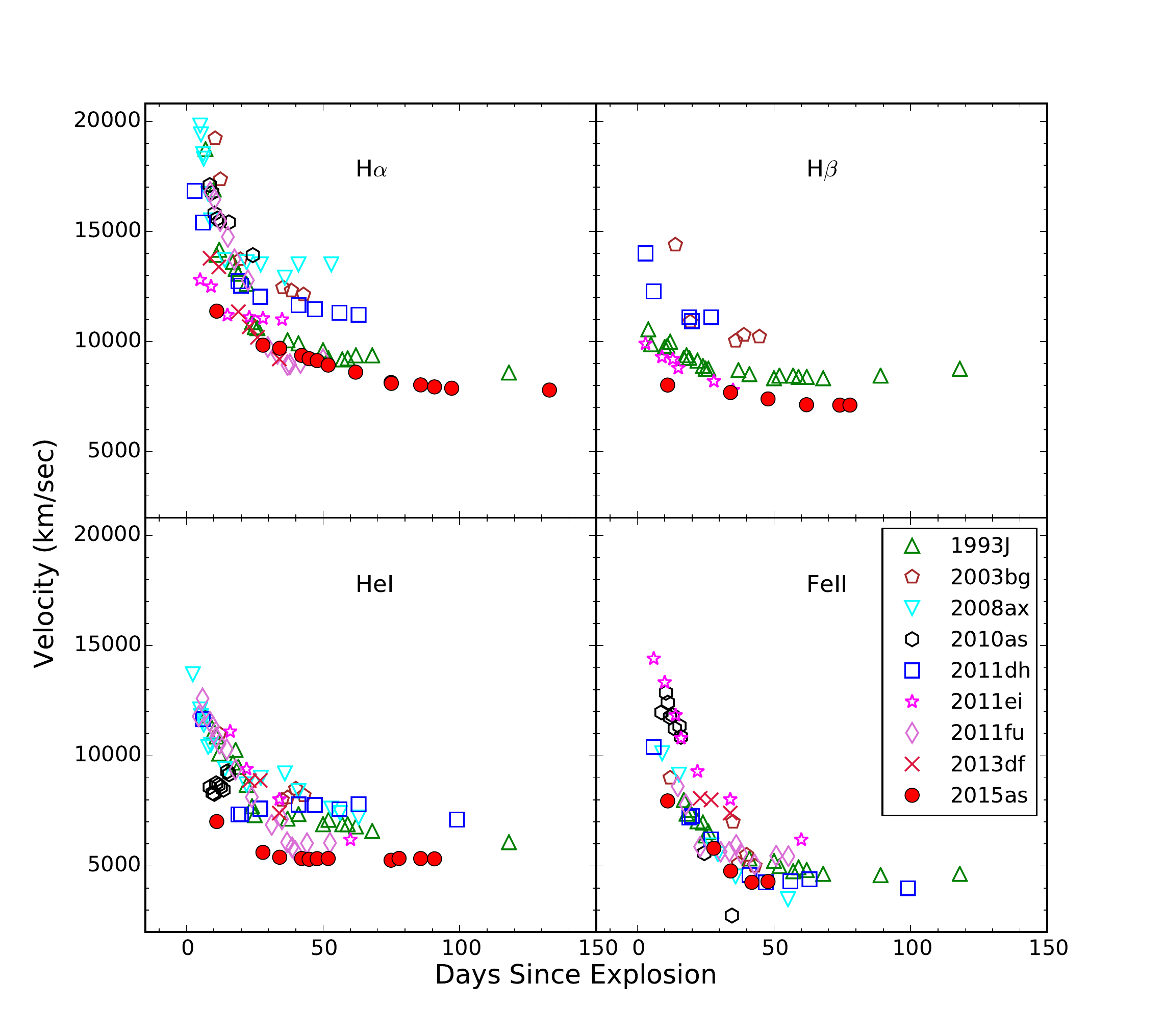} \\
	\end{center}
	\caption{Evolution of line expansion velocities of SN 2015as and a Type IIb SN sample. SN 2015as has the lowest velocity among the comparison sample.}
	\label{fig:vel_comp}
\end{figure}

In Fig. \ref{fig:vel_comp}, we compare the velocity evolution of SN 2015as with those of SNe 1993J, 2003bg, 2008ax, 2010as, 2011dh, 2011ei, 2011fu and 2013df. In the present Type IIb sample, SN 2015as has the lowest observed velocities. At about 10 days after explosion, the H$\alpha$ velocity for other SNe IIb like 2003bg, 2010as and 2011fu is 17,000  km sec$^{-1}$, while SNe 1993J, 2008ax, 2011dh and  2013df is $\sim$ 13,000 km sec$^{-1}$, which are about 17-18 $\%$ higher than that of SN 2015as, while SN 2011ei has a similar velocity of 12,500 km sec$^{-1}$ (see above; and top-left panel of Fig. \ref{fig:vel_comp}). Even though H$\beta$ and HeI starts with a lower velocity, a similar trend is also observed for the H$\beta$ (top-right in the figure) and the He I lines (bottom-left), while Fe II follows a behaviour similar to other SNe IIb (bottom-right). In SN 2015as, H$\alpha$ remained visible for a longer time at lower velocities than all the members of the comparison sample. \cite{1997ApJ...477..865I} showed that the minimum H velocity in SNe IIb mostly depends on the mass of the H envelope retained at the explosion time. For different SNe, the hydrogen mass retained is of the orders of 0.1 to 0.9 M$_{\odot}$ for SN 1993J, 0.06 M$_{\odot}$ for SN 2008ax and 0.1 M$_{\odot}$ for SN 2011dh \citep{2000AJ....120.1499M,2015ApJ...811..147F,2012ApJ...757...31B}. The SNe in the comparison sample have comparable H mass and similar explosion energy still SN 2015as has lower expansion velocity than other SNe. This indicates that in the case of SN 2015as, the SN ejecta has higher density. A significant fraction of the explosion energy of SN 2015as is used in expanding the ejecta, leading to lower velocity of the expanding ejecta. Indeed, our observations indicate a peak magnitude towards the faint end of the Type IIb SNe sample distribution, although it is brighter than the faintest SNe 2011ei and 2011hs.

\section{Conclusion}
\label{5}
In this paper, we present comprehensive {\it BVRIugriz} optical photometric and low-resolution spectroscopic monitoring of the Type IIb SN 2015as. Our photometric observing campaign started 3 days after the explosion and lasted over 500 days, while the spectroscopic observations lasted about 230 days. The colour evolution of SN 2015as have a remarkable resemblance to SN 2008ax, indicating that SN 2015as probably belongs to the subclass of SNe IIb without an early light curve peak. With a peak $V$ magnitude of -16.82, SN 2015as is brighter than SNe 1996cb, 2011ei and comparable with SN 2011hs. The early-time decay rate of SN 2015as in $B$-band closely matches that of SN 2011dh whereas it's slower than SN 2011fu. The $V$-band decline of SN 2015as is faster than SNe 2011dh and 2011fu, whereas the $R$-band and the $I$-band decay rates of SN 2015as are slower than SNe 2011dh and 2011fu and are faster than SN 1993J. The late-time light curve evolution of SN 2015as is slower than those of other SNe IIb.  
 
The comparison among quasi-bolometric light curves of SNe IIb shows that SN 2015as is one of the faintest events in our sample. In fact, the brightest object, SN 2011fu, has a peak luminosity of 5.2$\times$10$^{42}$ erg sec$^{-1}$, whereas SN 2015as has a maximum luminosity a factor 3.5 lower (1.5$\times$10$^{42}$ erg sec$^{-1}$). Using the formulation from \cite{1982ApJ...253..785A}, we estimate for SN 2015as M$_{Ni}$ = 0.08 M$_{\odot}$, E$_{k}$ = 0.65$\times$10$^{51}$ erg and $M_{ej}$ = 2.2  M$_{\odot}$. The semi-analytical model of \cite{2016A&A...589A..53N}, gives  M$_{Ni}$ = 0.08 M$_{\odot}$, E$_{k}$ = 1.0$\times$10$^{51}$ erg and $M_{ej}$ between 1.1 and 2.2  M$_{\odot}$. The $^{56}$Ni and ejecta masses of SN 2015as are similar to that of SN 2008ax, with a comparatively lower energy budget. The two-component model yields the progenitor radius of SN 2015as ($\sim$ 0.05$\times$10$^{13}$ cm), smaller than those of SN 1993J \citep[$\sim$ 1.9$\times$10$^{13}$ cm;][]{2012ApJ...757...31B}, SN 2011fu \citep[$\sim$ 0.3$\times$10$^{13}$ cm;][]{2015MNRAS.454...95M} but comparable with SN 2008ax  \citep[$\sim$ 0.03$\times$10$^{13}$ cm;][]{2015ApJ...811..147F}.

The early spectra of SN 2015as (until 75 days past explosion) show prominent Balmer lines, and are similar to those of SNe 1993J and 2011dh while the nebular phase spectra resemble those of SNe 1996cb and 2008ax. The photospheric velocity obtained from the pre-maximum synthetic spectrum, generated by SYN++ was found to be 8,500 km sec$^{-1}$. The photospheric velocities of the different lines in SN 2015as are found to be the lowest in the SN sample. The H$\alpha$ line was persistent for a long time in SN 2015as, which indicates that SN 2015as has more residual mass and a much slower dilution of the H envelope than other SNe IIb. Since the ejecta in SN 2015as has a higher density,  a significant fraction of the explosion energy is used in expanding the ejecta which results in lower expansion velocity (as seen in SN 2015as) and also a fainter primary peak.  The asymmetric profiles of [OI] 6300, 6364 \AA~ in the 132 days spectrum are well explained by the geometry of the emitting region. The minimum ejected mass of O ($\sim$ 0.45 M$_{\odot}$), estimated using the [OI]  6300, 6364 line flux, together with the information inferred from the [Ca II] 7291, 7324 / [OI] 6300, 6364 line ratio, suggest either a main sequence progenitor mass of $\sim$ 15 M$_{\odot}$ with a He core of 4 M$_{\odot}$ or a Wolf-Rayet star of 20-25 M$_{\odot}$ mass like SN 2003bg.

\section*{Acknowledgments}
We sincerely thank the anonymous referee for giving his/her valuable comments and suggestions and carefully going through the manuscript which has significantly improved the presentation of this work. We thank the observing staff and observing assistants at 104 cm ST, 130 cm DFOT, 182 cm EKAR Asiago Telescope and 200 cm HCT for their support during observations of SN 2015as. We acknowledge Wiezmann Interactive Supernova data REPository http://wiserep.weizmann.ac.il (WISeREP) \citep{2012PASP..124..668Y}. This research has made use of the CfA Supernova Archive, which is funded in part by the National Science Foundation through grant AST 0907903. AP, LT and SB are partially supported by the  PRIN-INAF 2014 with the project ``Transient Universe: Unveiling New Types of Stellar Explosions with PESSTO''. The work made use of the Swift Optical/Ultraviolet Supernova Archive (SOUSA).   SOUSA is supported by NASA's Astrophysics Data Analysis Program through grant NNX13AF35G.  BK acknowledges the Science and Engineering Research Board (SERB) under the Department of Science \& Technology, Govt. of India, for financial assistance in the
form of National Post-Doctoral Fellowship (Ref. no. PDF/2016/001563).

%---------------------FIGURES--------------------

%%%%%%%%%%%%%%%%%%%%%%%%%%%%%%%%%%%%%%%%%%%%%%%%%%

%%%%%%%%%%%%%%%%%%%% REFERENCES %%%%%%%%%%%%%%%%%%

% The best way to enter references is to use BibTeX:

%\bibliographystyle{mnras}
%\bibliography{example} % if your bibtex file is called example.bib

% Alternatively you could enter them by hand, like this:
% This method is tedious and prone to error if you have lots of references
%\begin{thebibliography}{99}
%\bibitem[\protect\citeauthoryear{Author}{2012}]{Author2012}
%Author A.~N., 2013, Journal of Improbable Astronomy, 1, 1
%\bibitem[\protect\citeauthoryear{Others}{2013}]{Others2013}
%Others S., 2012, Journal of Interesting Stuff, 17, 198
%\end{thebibliography}

%%%%%%%%%%%%%%%%%%%%%%%%%%%%%%%%%%%%%%%%%%%%%%%%%%

%%%%%%%%%%%%%%%%% APPENDICES %%%%%%%%%%%%%%%%%%%%%

%\appendix

%\section{Some extra material}

%If you want to present additional material which would interrupt the flow of the main paper,
%it can be placed in an Appendix which appears after the list of references.

%%%%%%%%%%%%%%%%%%%%%%%%%%%%%%%%%%%%%%%%%%%%%%%%%%

\bibliographystyle{mnras}
\bibliography{refag}
\begin{table*}
\caption{Star ID and the magnitudes in {\it BVRI} filters of the 15 secondary standards in the field of SN 2015as}
\centering
\smallskip
\begin{tabular}{c c c c c c c}
\hline \hline
Star ID  &$\alpha$      &$\delta$         &  $B$               &  $V$                  &  $R$                       &  $I$                       \\
         & (h:m:s)      &($^{\circ}$ ' '')& (mag)              & (mag)                 & (mag)                      & (mag)                     \\
\hline                                                                      
A	 & 10:07:38.619	& +51:54:47.56   & 17.91 $\pm$ 0.02  & 16.45 $\pm$ 0.01     & 15.24 $\pm$ 0.01          & 14.57 $\pm$ 0.01         \\
B	 & 10:08:05.918	& +51:54:39.08   & 19.02 $\pm$ 0.02  & 18.30 $\pm$ 0.01     & 17.61 $\pm$ 0.01          & 17.50 $\pm$ 0.02         \\
C	 & 10:07:46.080	& +51:54:20.63   & 17.49 $\pm$ 0.02  & 16.77 $\pm$ 0.01     & 16.07 $\pm$ 0.01          & 15.97 $\pm$ 0.01         \\ 
D	 & 10:08:21.902	& +51:54:02.51   & 15.29 $\pm$ 0.02  & 14.45 $\pm$ 0.01     & 13.68 $\pm$ 0.01          & 13.53 $\pm$ 0.01         \\
E	 & 10:08:21.899	& +51:54:03.94   & 17.16 $\pm$ 0.02  & 16.40 $\pm$ 0.01     & 15.68 $\pm$ 0.01          & 15.56 $\pm$ 0.01         \\
F	 & 10:08:22.652	& +51:48:37.21   & 14.69 $\pm$ 0.02  & 14.01 $\pm$ 0.01     & 13.33 $\pm$ 0.01          & 13.26 $\pm$ 0.01         \\
G	 & 10:08:36.587	& +51:48:10.04   & 18.53 $\pm$ 0.02  & 18.02 $\pm$ 0.01     & 17.40 $\pm$ 0.01          & 17.38 $\pm$ 0.02         \\
H	 & 10:08:24.906	& +51:47:54.81   & 15.18 $\pm$ 0.02  & 14.51 $\pm$ 0.01     & 13.84 $\pm$ 0.01          & 13.78 $\pm$ 0.01         \\
I	 & 10:08:33.907	& +51:47:26.63   & 15.85 $\pm$ 0.02  & 15.32 $\pm$ 0.01     & 14.70 $\pm$ 0.01          & 14.67 $\pm$ 0.01         \\
J	 & 10:08:07.421	& +51:46:21.31   & 17.11 $\pm$ 0.02  & 16.67 $\pm$ 0.01     & 16.10 $\pm$ 0.01          & 16.07 $\pm$ 0.01         \\
K	 & 10:08:16.957	& +51:46:21.78   & 19.63 $\pm$ 0.03  & 18.27 $\pm$ 0.01     & 17.13 $\pm$ 0.01          & 16.64 $\pm$ 0.01         \\
L	 & 10:08:32.515	& +51:46:19.18   & 17.27 $\pm$ 0.01  & 16.68 $\pm$ 0.01     & 16.03 $\pm$ 0.01          & 15.97 $\pm$ 0.01         \\
M        & 10:08:01.813	& +51:50:02.90   & 20.02 $\pm$ 0.05  & 18.74 $\pm$ 0.03     & 18.17 $\pm$ 0.04          & 17.61 $\pm$ 0.02         \\
N        & 10:08:25.579	& +51:50:53.46   & 20.36 $\pm$ 0.06  & 18.92 $\pm$ 0.03     & 18.07 $\pm$ 0.06          & 17.27 $\pm$ 0.04         \\
O        & 10:08:27.179	& +51:50:28.09   & 19.66 $\pm$ 0.05  & 19.14 $\pm$ 0.02     & 18.82 $\pm$ 0.05          & 18.54 $\pm$ 0.06          \\
\hline                                   
\end{tabular}
\label{tab:optical_observations}      
\end{table*}

\begin{table*}
\caption{Star ID and the magnitudes in the {\it ugriz} filters of 7 secondary standards (taken from SDSS DR12) in the field of SN 2015as}
\centering
\smallskip
\begin{tabular}{c c c c c c c c}
\hline \hline
Star ID  &$\alpha$      &$\delta$         &  $u$               &  $g$                  &  $r$                 &  $i$            &  $z$   \\
         & (h:m:s)      &($^{\circ}$ ' '')& (mag)              & (mag)                 & (mag)                & (mag)           & (mag)   \\
\hline                                                                      
D	 & 10:08:21.902	& +51:54:02.51    & 16.47$\pm$0.18     & 14.82$\pm$0.13        & 14.22$\pm$0.13       & 13.98$\pm$0.10  & 13.88$\pm$0.11 \\
E	 & 10:08:21.899	& +51:54:03.94    & 18.28$\pm$0.19     & 16.71$\pm$0.15        & 16.18$\pm$0.14       & 15.98$\pm$0.16  & 15.94$\pm$0.13   \\
F	 & 10:08:22.652	& +51:48:37.21    & 15.62$\pm$0.11     & 14.24$\pm$0.12        & 13.81$\pm$0.11       & 13.64$\pm$0.15  & 13.60$\pm$0.10   \\
G	 & 10:08:36.587	& +51:48:10.04    & 16.13$\pm$0.15     & 14.75$\pm$0.16        & 14.32$\pm$0.10       & 14.17$\pm$0.12  & 14.16$\pm$0.13    \\
M        & 10:08:01.813	& +51:50:02.90    & 20.81$\pm$0.24     & 18.93$\pm$0.20        & 18.05$\pm$0.20       & 17.68$\pm$0.19  & 17.38$\pm$0.15 \\
N        & 10:08:25.579	& +51:50:53.46    & 20.39$\pm$0.22     & 19.30$\pm$0.21        & 19.01$\pm$0.19       & 18.93$\pm$0.20  & 18.90$\pm$0.20  \\
O        & 10:08:27.179	& +51:50:28.09    & 22.39$\pm$0.28     & 19.71$\pm$0.23        & 18.36$\pm$0.18       & 17.77$\pm$0.18  & 17.48$\pm$0.15  \\
\hline                                   
\end{tabular}
\label{tab:optical_observations_SDSS}      
\end{table*}

\begin{center}
%\begin{longtable}{|c | c | c | c | c | c | c | c|}
\begin{table*}
\hspace{-0.5cm}
\caption{Log of optical observations}  		  
\begin{tabular}{c c c c c c c c c}
\hline \hline
Date       & JD             &  Phase$^\dagger$  & $u$               & $g$                & $r$               & $i$               &  $z$               & Telescope\\    
	   &                &  (Days)           & (mag)             & (mag)              & (mag)             & (mag)             & (mag)             &            \\
\hline
%\endfirsthead
2015/11/09.18 & 2457335.68     & 3.17   & ---                & ---                 & 17.49 $\pm$ 0.01 & ---              & ---              & EKAR \\
2015/11/17.18 & 2457343.68     & 11.18  & 16.65 $\pm$ 0.01   & ---  	           & 15.37 $\pm$ 0.01 & 15.67 $\pm$ 0.01 & 15.88 $\pm$ 0.02 & EKAR  \\
2015/12/03.16 & 2457360.64     & 28.14  & ---                & ---                 & 14.60 $\pm$ 0.02 & 14.67 $\pm$ 0.02 & ---              & EKAR \\
2015/12/10.08 & 2457366.56     & 34.06  & 17.18 $\pm$ 0.01   & 15.42 $\pm$ 0.01    & 14.71 $\pm$ 0.01 & 14.77 $\pm$ 0.01 & 15.01 $\pm$ 0.01 & EKAR \\
2015/12/17.21 & 2457374.69     & 42.20  & 18.13 $\pm$ 0.02   & 15.87 $\pm$ 0.01    & 14.97 $\pm$ 0.01 & 14.95 $\pm$ 0.01 & 15.07 $\pm$ 0.01 & EKAR  \\
2016/01/18.91 & 2457406.53     & 74.03  & ---	             & 16.78 $\pm$ 0.01    & 16.06 $\pm$ 0.01 & 15.88 $\pm$ 0.01 & 15.73 $\pm$ 0.01 & EKAR  \\
2016/01/19.93 & 2457407.43     & 74.93  & 18.60 $\pm$ 0.06   & 16.77 $\pm$ 0.01    & 16.04 $\pm$ 0.01 & 15.87 $\pm$ 0.01 & ---              & EKAR \\
2016/02/04.05 & 2457423.52     & 91.02  & 18.54 $\pm$ 0.04   & 16.91 $\pm$ 0.01    & 16.29 $\pm$ 0.01 & 16.17 $\pm$ 0.01 & 15.91 $\pm$ 0.01 & EKAR\\
2016/02/11.07 & 2457429.58     & 97.08  & 18.75 $\pm$ 0.03   & 16.96 $\pm$ 0.01    & 16.39 $\pm$ 0.01 & 16.29 $\pm$ 0.01 & 16.00 $\pm$ 0.01 & EKAR  \\
2016/03/17.92 & 2457465.39     & 132.89 & 18.80 $\pm$ 0.02   & 17.42 $\pm$ 0.01    & 16.85 $\pm$ 0.01 & 16.88 $\pm$ 0.01 & ---              & EKAR\\
2016/04/27.90 & 2457506.42     & 173.93 & ---                & 17.98 $\pm$ 0.02    & 17.35 $\pm$ 0.01 & 17.64 $\pm$ 0.01 & ---              & EKAR\\
2016/05/17.94 & 2457526.40     & 193.90 & ---                & 18.28 $\pm$ 0.03    & 17.64 $\pm$ 0.02 & 18.07 $\pm$ 0.03 & 17.79 $\pm$ 0.02 & EKAR\\
2016/06/23.92 & 2457563.39     & 230.89 & ---                & 18.42 $\pm$ 0.01    & 18.13 $\pm$ 0.01 & 18.94 $\pm$ 0.09 & 18.58 $\pm$ 0.01 & EKAR\\
2016/11/30.09 & 2457722.58     & 390.08 & ---                & 19.60 $\pm$ 0.03    & 19.79 $\pm$ 0.06 & 20.70 $\pm$ 0.01 & 20.53 $\pm$ 0.01 & EKAR \\
2017/03/28.96 & 2457841.44     & 508.94 & ---                & ---                 & 20.97 $\pm$ 0.12 & 21.68 $\pm$ 0.08 & ---              & EKAR \\
\hline\hline              
Date       & JD         & Phase$^\dagger$      &   $B$             &  $V$                   &     $R$                 &  $I$             & Telescope\\
           &            & (Days)               &  (mag)            &  (mag)                 &     (mag)               & (mag)            &          \\
\hline 
%\endhead
%\multicolumn{6}{c}        
2015/11/17.17 & 2457343.67 & 11.16      & 15.79 $\pm$ 0.01 &  15.52 $\pm$ 0.01    &     ---                 & ---                & EKAR \\
2015/12/01.94 & 2457358.43 & 25.93      & 14.92 $\pm$ 0.02 &  14.55 $\pm$ 0.04    &     13.97 $\pm$ 0.01    & 14.26 $\pm$ 0.01   & DFOT \\
2015/12/03.16 & 2457360.64 & 28.14      & 15.07 $\pm$ 0.01 &  14.65 $\pm$ 0.05    &     ---                 & ---                & EKAR \\
2016/12/05.95 & 2457362.42 & 29.91      & 15.24 $\pm$ 0.01 &  ---                 &     14.03 $\pm$ 0.01    & 14.30 $\pm$ 0.01   & DFOT  \\
2015/12/10.08 & 2457366.56 & 34.06      & 15.77 $\pm$ 0.01 &  14.95 $\pm$ 0.01    &     ---                 & ---                & EKAR \\
2015/12/17.24 & 2457374.69 & 42.18      & 16.46 $\pm$ 0.01 &  15.32 $\pm$ 0.01    &     ---                 & ---                & EKAR \\
2015/12/20.88 & 2457377.38 & 44.87      & 16.47 $\pm$ 0.01 &  15.54 $\pm$ 0.01    &     14.65 $\pm$ 0.01    & 14.68 $\pm$ 0.01   & DFOT \\ 
2015/12/23.83 & 2457380.31 & 47.81      & 16.78 $\pm$ 0.01 &  15.58 $\pm$ 0.01    &     14.70 $\pm$ 0.02    & 14.70 $\pm$ 0.01   & HCT \\
2016/01/01.85 & 2457389.35 & 56.85      & ---              &  15.87 $\pm$ 0.01    &     15.09 $\pm$ 0.01    & 15.02 $\pm$ 0.01   & ST \\
2016/01/02.85 & 2457390.35 & 57.85      & ---              &  15.90 $\pm$ 0.01    &     15.13 $\pm$ 0.01    & 15.07 $\pm$ 0.01   & ST \\
2016/01/03.93 & 2457391.41 & 58.91      & ---              &  15.92 $\pm$ 0.02    &     ---                 & 15.06 $\pm$ 0.01   & ST \\
2016/01/05.83 & 2457393.31 & 60.81      & ---              &  15.96 $\pm$ 0.02    &     15.20 $\pm$ 0.01    & 15.14 $\pm$ 0.01   & ST \\
2016/01/12.83 & 2457400.36 & 67.85      & ---              &  16.11 $\pm$ 0.04    &     ---                 & 15.24 $\pm$ 0.02   & ST \\
2016/01/13.84 & 2457401.33 & 68.83      & ---              &  16.11 $\pm$ 0.04    &     15.33 $\pm$ 0.01    & 15.25 $\pm$ 0.01   & ST \\
2016/01/15.89 & 2457403.38 & 70.87      & ---              &  16.15 $\pm$ 0.05    &     15.36 $\pm$ 0.01    & 15.27 $\pm$ 0.01   & DFOT \\
2016/01/16.96 & 2457404.44 & 71.93      & ---              &  16.22 $\pm$ 0.02    &     15.38 $\pm$ 0.01    & 15.24 $\pm$ 0.01   & DFOT \\
2016/01/18.05 & 2457406.52 & 74.02      & 17.02 $\pm$ 0.02 &  16.27 $\pm$ 0.01    &     ---                 & ---                & EKAR \\
2016/01/19.06 & 2457407.42 & 74.91      & 17.06 $\pm$ 0.01 &  16.26 $\pm$ 0.01    &     ---                 & ---                & EKAR \\
2016/01/20.94 & 2457408.43 & 75.93      & ---              &  16.27 $\pm$ 0.04    &     ---                 & ---                & ST \\
2016/01/22.88 & 2457410.36 & 77.85      & 17.09 $\pm$ 0.01 &  16.28 $\pm$ 0.01    &     15.47 $\pm$ 0.01    & 15.49 $\pm$ 0.01   & HCT \\
2016/01/25.83 & 2457413.31 & 80.81      & 17.12 $\pm$ 0.01 &  16.32 $\pm$ 0.01    &     15.54 $\pm$ 0.02    & 15.51 $\pm$ 0.01   & HCT \\
2016/01/30.80 & 2457418.27 & 85.77      & ---              &  16.44 $\pm$ 0.02    &     15.63 $\pm$ 0.01    & 15.52 $\pm$ 0.01   & ST \\
2016/01/31.78 & 2457419.28 & 86.77      & 17.19 $\pm$ 0.02 &  16.45 $\pm$ 0.02    &     15.65 $\pm$ 0.01    & 15.52 $\pm$ 0.01   & ST \\
2016/02/01.81 & 2457420.29 & 87.79      & ---              &  16.46 $\pm$ 0.02    &     ---                 & 15.57 $\pm$ 0.01   & ST \\
2016/02/02.75 & 2457421.22 & 88.72      & ---              &  16.47 $\pm$ 0.02    &     15.70 $\pm$ 0.01    & 15.63 $\pm$ 0.01   & ST \\
2016/02/05.05 & 2457423.52 & 91.02      & 17.23 $\pm$ 0.01 &  16.48 $\pm$ 0.01    &     ---                 & ---                & EKAR \\
2016/02/10.75 & 2457429.25 & 96.75      & 17.31 $\pm$ 0.02 &  16.54 $\pm$ 0.01    &     15.79 $\pm$ 0.01    & 15.64 $\pm$ 0.01   & ST, EKAR \\ 
2016/02/11.86 & 2457430.34 & 97.83      & ---              &  ---       	  &     15.82 $\pm$ 0.01    & 15.70 $\pm$ 0.02   & ST \\
2016/02/12.83 & 2457431.31 & 98.81      & 17.34 $\pm$ 0.02 &  16.62 $\pm$ 0.03    &     ---                 & 15.70 $\pm$ 0.03   & DFOT \\
2016/02/13.88 & 2457432.36 & 99.85      & 17.35 $\pm$ 0.02 &  16.65 $\pm$ 0.01    &     15.86 $\pm$ 0.02    & 15.72 $\pm$ 0.01   & DFOT \\
2016/02/14.90 & 2457433.37 & 100.87     & ---              &  16.68 $\pm$ 0.03    &     15.85 $\pm$ 0.01    & ---                & ST \\
2016/02/24.71 & 2457443.19 & 110.68     & 17.46 $\pm$ 0.01 &  16.73 $\pm$ 0.01    &     15.90 $\pm$ 0.01    & 15.86 $\pm$ 0.01   & HCT \\
2016/02/27.75 & 2457446.25 & 113.75     & ---              &  ---                 &     16.01 $\pm$ 0.01    & 15.91 $\pm$ 0.01   & ST \\
2016/02/28.84 & 2457447.30 & 114.79     & 17.50 $\pm$ 0.01 &  16.76 $\pm$ 0.01    &     16.03 $\pm$ 0.01    & 15.94 $\pm$ 0.01   & HCT \\
2016/03/01.81 & 2457448.29 & 115.79     & ---              &  ---                 &     16.06 $\pm$ 0.01    & 15.95 $\pm$ 0.01   & ST \\ 
2016/03/02.75 & 2457450.25 & 117.75     & ---       	   &  ---      	          &     16.09 $\pm$ 0.01    & ---                & ST \\
2016/03/17.91 & 2457465.39 & 132.89     & 17.74 $\pm$ 0.01 &  17.06 $\pm$ 0.01    &     ---                 & ---                & EKAR \\
2016/03/18.83 & 2457466.31 & 133.81     & ---	           &  --- 	          &     16.32 $\pm$ 0.01    & 16.22 $\pm$ 0.01   & ST \\
2016/03/28.76 & 2457476.24 & 143.74     & --- 	           &  ---      	          &     16.46 $\pm$ 0.01    & 16.47 $\pm$ 0.02   & ST \\
2016/04/08.66 & 2457487.14 & 154.64     & ---              &  17.60 $\pm$ 0.03    &     16.55 $\pm$ 0.02    & 16.64 $\pm$ 0.03   & DFOT \\
2016/04/12.74 & 2457491.23 & 158.73     & 17.95 $\pm$ 0.01 &  ---                 &     ---                 & ---                & HCT \\
2016/04/20.75 & 2457499.22 & 166.72     & 18.11 $\pm$ 0.02 &  17.88 $\pm$ 0.01    &     16.63 $\pm$ 0.01    & 16.79 $\pm$ 0.02   & HCT \\
2016/04/27.94 & 2457506.43 & 173.93     & 18.29 $\pm$ 0.02 &  17.95 $\pm$ 0.01    &     16.75 $\pm$ 0.02    & 16.89 $\pm$ 0.01   & HCT, EKAR \\
2016/05/17.97 & 2457526.40 & 193.89     & 18.58 $\pm$ 0.05 &  18.32 $\pm$ 0.02    &     ---                 & ---                & EKAR \\
2016/06/30.68 & 2457570.18 & 237.68     & ---              &  19.10 $\pm$ 0.01    &     17.34 $\pm$ 0.02    & ---                & HCT \\
2016/11/30.10 & 2457722.57 & 390.06     & ---              &  21.40 $\pm$ 0.01    &     ---                 & ---                & EKAR \\
2016/12/06.18 & 2457728.68 & 396.18     & ---              &  21.50 $\pm$ 0.01    &     ---                 & ---                & EKAR \\ 
\hline          		                    
\end{tabular}
\label{tab:observation_log}     
%\hline
$^\dagger$ Phase has been calculated since explosion JD = 2457332.5 
\end{table*}
%\end{longtable}
\end{center}

\begin{table*}
\caption{Log of UV observations}
\centering
\smallskip
\begin{tabular}{c c c c c c c}
\hline \hline
Phase      	   & $uvw2$               & $uvw1$              & $uvm2$            &  $u$              &  $b$                  & $v$  \\
(JD)     	   &(mag)                 &(mag)                &(mag)              & (mag)             & (mag)                 & (mag)  \\
\hline
2457342.82 	   & 18.21 $\pm$ 0.18    & 17.33 $\pm$ 0.13	& 19.23 $\pm$ 0.27 & 15.75 $\pm$ 0.07	& 15.83 $\pm$ 0.06	& 15.48 $\pm$ 0.07 \\
2457344.62	   & 18.41 $\pm$ 0.22    & 17.17 $\pm$ 0.13	& 18.95 $\pm$ 0.31 & 15.73 $\pm$ 0.08	& 15.54 $\pm$ 0.06	& 15.17 $\pm$ 0.07 \\
2457346.05	   & 17.94 $\pm$ 0.13    & 16.98 $\pm$ 0.09	& 19.14 $\pm$ 0.29 & 15.30 $\pm$ 0.06	& 15.33 $\pm$ 0.05	& 15.09 $\pm$ 0.06 \\
2457348.84	   & 17.78 $\pm$ 0.13    & 16.80 $\pm$ 0.09	& 19.06 $\pm$ 0.26 & 14.87 $\pm$ 0.05	& 15.06 $\pm$ 0.05	& 14.84 $\pm$ 0.06 \\
\hline                                   
\end{tabular}
\label{tab:UV_observations}      
\end{table*}

\begin{center}
\begin{table*}
\caption{Log of spectroscopic observations}  		  
\begin{tabular}{c c c c c c c c}
\hline \hline
Date          & Phase$^\dagger$& Grism      & Spectral Range       & Resolution   & Exposure Time & Slit Width  & Telescope                \\
(UT)          &(Days)          &            & (\AA~)               &              & (sec)         & (arcsecs)            &                 \\
\hline
2015/11/17.17 & 11.1           & Gr04       & 3360-7740            & 311          & 1800   	& 1.26	        & AFOSC, Ekar \\
2015/12/03.23 & 27.9           & Gr07,Gr08  & 3800-6840,5800-8350  & 1330,2190    &  900   	& 0.77, 1.92	& HFOSC, HCT    \\
2015/12/05.82 & 29.8           & Gr07,Gr08  & 3800-6840,5800-8350  & 1330,2190    &  900	& 0.77, 1.92	& HFOSC, HCT    \\
2015/12/10.88 & 34.1           & VPH6,VPH7  & 4500-10000,3200-7000 & 500,470      & 1500   	& 1.69, 2.50    & AFOSC, Ekar  \\
2015/12/17.82 & 42.1           & VPH6,VPH7  & 4500-10000,3200-7000 & 500,470      & 1200   	& 1.69, 2.50    & AFOSC, Ekar \\
2015/12/20.73 & 44.9           & Gr07,Gr08  & 3800-6840,5800-8350  & 1330,2190    & 1800        & 0.77, 1.92	& HFOSC, HCT     \\
2015/12/23.45 & 47.8           & Gr07,Gr08  & 3800-6840,5800-8350  & 1330,2190    &  600        & 0.77, 1.92	& HFOSC, HCT      \\
2015/12/27.32 & 51.8           & Gr07,Gr08  & 3800-6840,5800-8350  & 1330,2190    & 2400        & 0.77, 1.92	& HFOSC, HCT       \\
2015/01/06.56 & 61.9           & Gr07,Gr08  & 3800-6840,5800-8350  & 1330,2190    & 1800        & 0.77, 1.92	& HFOSC, HCT          \\
2016/01/18.21 & 74.0           & VPH6,VPH7  & 4500-10000,3200-7000 & 500,470      &  300   	& 1.69, 2.50    & AFOSC, Ekar  \\
2016/01/19.14 & 74.8           & VPH6,VPH7  & 4500-10000,3200-7000 & 500,470      & 1800   	& 1.69, 2.50    & AFOSC, Ekar        \\
2016/01/20.05 & 75.0           & Gr07,Gr08  & 3800-6840,5800-8350  & 1330,2190    &  300        & 0.77, 1.92	& HFOSC, HCT     \\
2016/01/22.08 & 77.8           & Gr07,Gr08  & 3800-6840,5800-8350  & 1330,2190    & 1500        & 0.77, 1.92	& HFOSC, HCT        \\
2016/01/30.81 & 85.7	       & Gr07,Gr08  & 3800-6840,5800-8350  & 1330,2190    & 2100        & 0.77, 1.92	& HFOSC, HCT     \\
2016/02/04.22 & 90.7           & Gr07,Gr08  & 3800-6840,5800-8350  & 1330,2190    & 2100       	& 0.77, 1.92	& HFOSC, HCT     \\
2016/02/04.12 & 90.9           & VPH6,VPH7  & 4500-10000,3200-7000 & 500,470      & 1200   	& 1.69, 2.50    & AFOSC, Ekar     \\
2016/02/10.19 & 97.0           & VPH6,VPH7  & 4500-10000,3200-7000 & 500,470      & 1500        & 1.69, 2.50	& AFOSC, Ekar \\
2016/02/20.33 & 106.7	       & Gr07,Gr08  & 3800-6840,5800-8350  & 1330,2190    & 1800        & 0.77, 1.92	& HFOSC, HCT    \\
2016/02/24.12 & 110.8	       & Gr07,Gr08  & 3800-6840,5800-8350  & 1330,2190    & 2400        & 0.77, 1.92	& HFOSC, HCT    \\
2016/03/17.09 & 132.8          & VPH6,VPH7  & 4500-10000,3200-7000 & 500,470      & 1800        & 1.69, 2.50	& AFOSC, Ekar      \\
2016/03/31.02 & 146.7          & Gr07,Gr08  & 3800-6840,5800-8350  & 1330,2190    & 2100        & 0.77, 1.92	& HFOSC, HCT  \\
2016/04/27.01 & 173.8          & VPH6       & 4500-10000           & 500,470      & 2400        & 1.69		& AFOSC, Ekar  \\
2016/04/28.08 & 174.6          & Gr07,Gr08  & 3800-6840,5800-8350  & 1330,2190    & 1200        & 0.77, 1.92	& HFOSC, HCT  \\
2016/06/23.66 & 230.8          & Gr04       & 3360-7740            & 311          & 2400        & 1.26		& AFOSC, Ekar   \\
\hline                                   
\end{tabular}
\label{tab:spec_observations}     
\newline
$^\dagger$ Phase has been calculated since explosion JD = 2457332.5 
\end{table*}
%\end{longtable}
\end{center}

\begin{table*}
\caption{Light curve parameters of SN 2015as}
\centering
\smallskip
\begin{tabular}{c c c c c}
\hline \hline
Band         	          & Peak Observed mag   & Peak abs. mag.          & Rise Time       \\
       	     	          & (mag)               & (mag)                   & (Days)           \\
\hline
$B$		          & 14.74 $\pm$ 0.02   & -16.65 $\pm$ 0.39       & 20.88 $\pm$ 0.30     \\    
$V$		          & 14.63 $\pm$ 0.03   & -16.82 $\pm$ 0.18       & 24.81 $\pm$ 0.79	\\
$r$	                  & 14.48 $\pm$ 0.01   & -17.01 $\pm$ 0.39       & 23.53 $\pm$ 0.48	\\
$i$		          & 14.68 $\pm$ 0.01   & -16.68 $\pm$ 0.38       & 26.46 $\pm$ 0.49	\\
\hline                                   
\end{tabular}
\label{tab:photometric_properties}      
\end{table*}										

\begin{table*}
\caption{Properties of the comparison sample }
\centering
\smallskip
%\footnote
\begin{tabular}{c c c c c c c c c c}
\hline \hline
                     & SNe         & Distance      & Extinction &  M$_V$          &E$_{k}$  	   &$\triangle m_{15}$($V$)&$^{56}$Ni mass& M$_{ej}$         & Reference$^\dagger$  \\
                     &             & (Mpc)         &$E(B-V)$    &  (mag)          &(10$^{51}$ erg) & (mag)                 &(M$_{\odot}$) &(M$_{\odot}$)        &     \\
\hline
Type IIb             & SN 1993J    & 3.63$\pm$0.7  & 0.180      & -17.59$\pm$0.13 & 0.7-1.4        & 0.98$\pm$0.04  & 0.10-0.14     & 1.3-3.5         & 1,11 \\
                     & SN 1996cb   & 6.41          & 0.030      & -16.22	  & --             & 0.97$\pm$0.04  & --            & --      	       & 2,11,12 \\
                     & SN 2003bg   & 21.7          & 0.024      & -16.95	  & 5              & 0.96$\pm$0.03  & 0.1-0.2       & 4	      
& 3,11,12  \\
                     & SN 2008ax   & 9.6$\pm$1.3   & 0.422      & -17.61$\pm$0.43 & 1-6            & 0.91$\pm$0.03  & 0.07-0.15     & 2-5     	       & 4,11   \\
                     & SN 2010as   & 27.3$\pm$4.7  & 0.820      & -18.01	  & 0.7            & 0.94$\pm$0.05  & 0.12          & 2.5             & 5,11,12   \\
                     & SN 2011dh   & 8.4$\pm$0.7   & 0.035      & -17.12$\pm$0.18 & 0.6-1.0        & 0.98$\pm$0.04  & 0.05-0.10     & 1.8-2.5         & 6,11    \\   
                     & SN 2011ei   & 28.5$\pm$5.7  & 0.240      & -16.00          & 2.5            & 0.75$\pm$0.13  & 0.03          & 1.6
& 13      \\
                     & SN 2011hs   & 18.3$\pm$0.1  & 0.170      & -16.59	  & 0.85           & 1.18$\pm$0.06  & 0.04	    & 1.8-2.5 	       & 7,11,12   \\
                     & SN 2011fu   & 77.9$\pm$5.5  & 0.218      & -18.50$\pm$0.24 & 1.3            & 0.78$\pm$0.04  & 0.15          & 3-5             & 8,11,12   \\
                     & SN 2013df   & 16.6$\pm$0.4  & 0.090      & -16.85$\pm$0.08 & 0.4-1.2        & 1.32$\pm$0.10  & 0.10-0.13     & 0.8-1.4         & 9,11,12   \\
\hline
Type Ib/Ic           &             &               &            & -18.07$\pm$0.06 &                &                &               &                 & 10     \\ 
\hline                                                                                   
\end{tabular}
\newline
$^\dagger$ REFERENCES.--(1) \cite{1995A&AS..110..513B}, \cite{1994AJ....107.1022R}; (2) \cite{1999AJ....117..736Q}; (3) \cite{2009ApJ...703.1612H}, NED; (4) \cite{2011MNRAS.413.2140T}, \cite{2008MNRAS.389..955P}; (5) \cite{2014ApJ...792....7F}; (6) \cite{2013MNRAS.433....2S}; (7) \cite{2013ApJ...767...71M}; (8) \cite{2014MNRAS.439.1807B}; (9) \cite{2013MNRAS.431..308K};  (10) \cite{2014MNRAS.445.1647M}; (11) \cite{2016PhDT.......113M}; (12) \cite{2011ApJ...741...97D}; (13) This work
\label{tab:photometric_parameters_different_SNe}      
\end{table*}

\begin{table*}
\caption{Rise times in {\it BVri} bands for a sample of Type IIb SNe.}
\centering
\smallskip
\begin{tabular}{c c c c c}
\hline \hline
Supernova           &                            & Rise Time (Days)                                                        \\    
\hline
                    &          $B$               &  $V$             &  $r$              & $i$              \\  
\hline
SN 1993J	    &          8.8$\pm$3.3       & 8.9$\pm$1.4      & 8.8$\pm$1.0       & 9.1$\pm$1.5           \\
SN 2008ax           &          18.9              & 20.7             & 22.3              & 22.8               \\
SN 2011dh           &          19.6$\pm$0.5      & 20.6$\pm$0.5     & 21.3$\pm$0.5      & 22.9$\pm$0.5   \\   
SN 2011fu           &          13.7$\pm$1.4      & 12.7$\pm$1.6     & 12.9$\pm$1.8      & 13.5$\pm$1.8    \\
SN 2015as	    &          20.8$\pm$0.3      & 24.8$\pm$0.8     & 23.5$\pm$0.5      & 26.4$\pm$0.5    \\
\hline                                   
\end{tabular}
\label{tab:rise times}      
\end{table*}											

\begin{table*}
\caption{Early and late time decay rates of a sample of Type IIb SNe. The sample includes four Type IIb SNe 1993J, 2008ax, 2011dh and 2011fu.}
\centering
\smallskip
\begin{tabular}{c c c c c c c}
\hline \hline
Supernova           &                          & Early time (50-100 days) decay rates   \\        
                    &                          & in units of mag (100 days)$^{-1}$   \\
\hline
                    &          $B$            & $V$              & $R$               & $I$              & Reference$^\dagger$  \\  
\hline
SN 1993J            &          1.46           & 1.73             & 1.57              & 1.77             & 1,2\\
SN 1996cb           &          1.23$\pm$0.09  & 1.96$\pm$0.05    & 2.15$\pm$0.06     & --               & 6  \\
SN 2008ax           &          1.46$\pm$0.02  & 2.05$\pm$0.06    & 2.20$\pm$0.07     & 1.90$\pm$0.07    & 6  \\
SN 2010as           &          0.80$\pm$0.10  & 1.67$\pm$0.04    & 2.30$\pm$0.12     & 1.96$\pm$0.20    & 6  \\
SN 2011dh           &          1.09$\pm$0.15  & 1.76$\pm$0.04    & 2.16$\pm$0.05     & 1.90$\pm$0.05    & 2  \\   
SN 2011fu           &          1.25$\pm$0.07  & 1.78$\pm$0.04    & 2.04$\pm$0.04     & 1.97$\pm$0.05    & 5 \\
SN 2013df           &          1.04$\pm$0.04  & 2.01$\pm$0.04    & --                & --               & 6 \\
SN 2015as	    &          1.10$\pm$0.03  & 1.85$\pm$0.06    & 1.87$\pm$0.08     & 1.73$\pm$0.09    & 6  \\
\hline
Supernova           &                  & Late time (100-300 days) decay rates           \\
                    &                  & in units of mag (100 days)$^{-1}$               \\
\hline                                                                                          
                    & $B$             & $V$                & $R$               & $I$               & Reference$^\dagger$ \\
\hline                                                                                                            
SN 1993J	    & 1.39            & 1.71               & 1.49              & 1.87              & 1,2 \\
SN 1996cb           & --              & 1.02$\pm$0.12      & 1.17$\pm$0.15     & --                & 6 \\       
SN 2008ax	    & 1.74$\pm$0.30   & 1.90$\pm$0.10      & 1.64$\pm$0.90     & 2.03$\pm$0.80     & 3,4 \\
SN 2011dh           & 1.71$\pm$0.13   & 1.83$\pm$0.11      & 1.51$\pm$0.05     & 1.70$\pm$0.06     & 2 \\ 
SN 2011hs           & 1.95$\pm$0.12   & 2.30$\pm$0.66      & 2.03$\pm$0.09     & --                & 6  \\
SN 2011fu           & --              & --                 & --                & --                & --   \\
SN 2015as	    & 1.30$\pm$0.08   & 1.65$\pm$0.04      & 1.13$\pm$0.04     & 1.67$\pm$0.03     & 6  \\ 
\hline  
\end{tabular}
\newline
$^\dagger$ REFERENCES.--(1) \cite{1995A&AS..110..513B}; (2) \cite{2013MNRAS.433....2S}; (3) \cite{2008MNRAS.389..955P}; (4) \cite{2011MNRAS.413.2140T}; (5) \cite{2015MNRAS.454...95M};  (6) This work
\label{tab:comp_decay_rate} 
\end{table*}

\begin{table*}
\caption{Best fit parameters derived from the analytical modelling of the bolometric light curve using \citet{2016A&A...589A..53N}.}
\centering
\smallskip
\begin{tabular}{c c c c c c c}
\hline \hline 
Pseudo-bolometric lightcurve parameters \\
\hline
Parameter                       &  Core(He-rich)		      &	Shell(H-rich)                      & Remarks   \\
	              		&  ($\kappa$ =0.24 cm$^{2}$ g$^{-1}$) &	($\kappa$ =0.4 cm$^{2}$ g$^{-1}$)  &           \\
\hline
{\it R$_{0}$} (cm)   		&  2 x 10$^{11}$		      & 0.05 x 10$^{13}$		   & Initial radius of the ejecta \\
{\it T$_{rec}$} (K)   		&  5500				      & --				   & Recombination Temperature \\
{\it M$_{ej}$} (M$_{\odot}$)	&  1.0				      & 0.1				   & Ejecta Mass \\
{\it M$_{Ni}$} (M$_{\odot}$)	&  0.08				      & --				   & Initial Nickel Mass \\
{\it E$_{Th}$}                  &  0.36	x 10$^{51}$	              & 0.30 x 10$^{51}$	           & Initial Thermal Energy \\
{\it E$_{kin}$}                 &  0.75	x 10$^{51}$		      & 0.25 x 10$^{51}$                   & Initial Kinetic Energy \\
\hline  
Full-bolometric light curve parameters   \\
\hline
Parameter                       &  Core(He-rich)		      &	Shell(H-rich)                      & Remarks   \\
	              		&  ($\kappa$ =0.24 cm$^{2}$ g$^{-1}$) &	($\kappa$ =0.4 cm$^{2}$ g$^{-1}$)  &           \\
\hline
{\it R$_{0}$} (cm)   		&  2 x 10$^{11}$		      & 0.05 x 10$^{13}$		   & Initial radius of the ejecta \\
{\it T$_{rec}$} (K)   		&  5500				      & --				   & Recombination Temperature \\
{\it M$_{ej}$} (M$_{\odot}$)	&  1.2				      & 0.1				   & Ejecta Mass \\
{\it M$_{Ni}$} (M$_{\odot}$)	&  0.08				      & --				   & Initial Nickel Mass \\
{\it E$_{Th}$}                  &  0.36	x 10$^{51}$	              & 0.30 x 10$^{51}$	           & Initial Thermal Energy \\
{\it E$_{kin}$}                 &  0.78	x 10$^{51}$		      & 0.28 x 10$^{51}$                   & Initial Kinetic Energy \\
\hline                              
\end{tabular}
\label{tab:Nagy}      
\end{table*}

% Don't change these lines
\bsp	% typesetting comment
\label{lastpage}
\end{document}